\documentclass[superscriptaddress,amsmath,amssymb,preprint,pre]{revtex4-1}
\usepackage{graphicx}
\usepackage{dcolumn}
\usepackage{bm}
\usepackage{color}
\usepackage[lofdepth,lotdepth,caption=false]{subfig}
\begin{document}
\title{Ensemble Distribution for Immiscible Two-Phase Flow in Porous Media}
\author{Isha Savani}
\email{Isha.Savani@gmail.com}
\affiliation{Department of Physics,
Norwegian University of Science and Technology, NTNU, N-7491 Trondheim, Norway}
\author{Dick Bedeaux}
\email{Dick.Bedeaux@chem.ntnu.no}
\affiliation{Department of Chemistry, Norwegian University of Science and 
Technology,  NTNU, N-7491 Trondheim, Norway}
\author{Signe Kjelstrup}
\email{Signe.Kjelstrup@ntnu.no}
\affiliation{Department of Chemistry, Norwegian University of Science and 
Technology,  NTNU, N-7491 Trondheim, Norway}
\author{Morten Vassvik}
\email{Morten.Vassvik@ntnu.no}
\affiliation{Department of Physics,  Norwegian University of Science and 
Technology,  NTNU, N-7491 Trondheim, Norway}
\author{Santanu Sinha}
\email{Santanu@csrc.ac.cn}
\affiliation{Beijing Computational Science Research Center, 
10 East Xibeiwang Road, Haidian, Beijing 100193, China}
\author{Alex Hansen}
\email{Alex.Hansen@ntnu.no}
\affiliation{Department of Physics,  Norwegian University of Science and 
Technology,  NTNU, N-7491 Trondheim, Norway}
\date{\today}
\begin{abstract}
We construct an ensemble distribution to describe steady immiscible two-phase 
flow of two incompressible fluids in a porous medium. The system is found to 
be ergodic. The distribution is used to compute macroscopic flow parameters. 
In particular, we find an expression for the overall mobility 
of the system from the ensemble distribution. The entropy production 
at the scale of the porous medium is shown to give the expected  product of 
the average flow and its driving force, obtained from a black-box 
description.  We test numerically some of the central theoretical results. 
\end{abstract}
\pacs{47.56.+r, 47.55.Ca, 47.55.dd, 89.75.Fb}
\maketitle

\section{Introduction}
\label{intro} 

Multiphase flow in porous media poses interesting problems to
engineers and scientists in diverse fields \cite{b88}. Understanding
the nature of multiphase flow is relevant to understand the flow of
particles in bifurcating blood vessels, or to categorizing liquid
transportation through cellulose. Other interesting areas concern
diffusion of pollutants in soil, and the flow of hydrocarbons and
water in oil reservoirs.

It is apparent from this wide range of applications of porous flow,
that the length scale of relevant processes can range from few
nanometers to several kilometers.  In geological transport processes
such as aquifers and oil reservoirs, this fact is especially
important, as the processes that occur at the pore scale (micron
scale) remain important in attempting to understand the processes at
the reservoir scale (kilometer scale).

When two immiscible fluids flow simultaneously in a rigid porous
medium, the state-of-the-art description is given by the relative
permeability equations, which are considered to be the effective
medium equations. The relative permeability approach views each fluid
as moving in a pore space that is constrained by the other fluid.
Hence, each fluid will experience a lowered effective permeability
since it experiences a diminished pore space in which to move.  The
ratio between the effective permeabilities of each fluid and the
single-fluid permeability of the porous medium are the relative
permeabilities.  The relative permeabilities are thought only to
depend on the fluid saturations (which are the volumes of each fluid
relative to the pore volume).  In addition to the relative
permeabilities, a capillary pressure field that models the interfacial
tension between the two fluids is introduced \cite{wb36}.

The concept of relative permeability is simple. However, different
laboratory methods, e.g.\ the Penn State or the Hassler method
\cite{orkhb51} yield different results for the measurement of relative
permeability. This signals that the relative permeability equations do
not offer a complete description of the problem.  These weaknesses
have been known for a long time and it is not controversial to state
that the relative permeability approach should and probably will be
replaced by a better framework.  Several attempts have been made to
replace this framework, see e.g.,
\cite{lsd81,g89,hg90,hg93a,hg93b,gh98,h98,g99,hb00,h06a,h06b,h06c,%
  hr09,hd10,nbh11,dhh12,habgo15,h15,vd16,gsd16,hsbksv16}.

Techniques for recording and reconstructing the pore structure of
porous media has developed tremendously over the last years
\cite{bbdgimpp13}.  It is now possible to render detailed maps of the
structure of porous media at the sub-pore level.

Numerical techniques to calculate the flow properties have also
developed and branched over the years.  There are several approaches.
Bryant and Blunt \cite{bb92} were the first to calculate relative
permeabilities from a detailed network model.  Aker et
al.\ \cite{amhb98,kah02} developed a network model which was extended
to include film flow by T{\o}r{\aa} et al.\ \cite{toh12}. The model is
today being combined with a Monte Carlo technique \cite{sshbkv16} to
speed up the calculations considerably. A recent review summarizes the
status of this class of models, see \cite{jh12}. A very different
approach is the Lattice Boltzmann method \cite{rob10,rino12}, see also
\cite{broglekawbw16,acbrsb16}.  Whereas the network models are ideal
for large network without detailed knowledge of the presice shape of
each pore, the Lattice Boltzmann method has the opposite strength.  It
goes well with the detailed pore spaces that are now being
reconstructed but is less useful in large networks.  Other methods
than the Lattice Boltzmann one which resolve the flow at the pore
level are e.g.\ smoothed particle hydrodynamics \cite{tm05,op10,ll10},
and density functional hydrodynamics \cite{abdekks16}.

The goal of any theory of immiscible two-phase flow in porous media must
be to bind together the physics at the pore level with a description
at scales where the porous medium may be seen as a continuum.  We illustrate
this viewpoint through the relative permeability equations that attempt to do 
exactly this:
\begin{equation}
\label{eqn7}
\vec \nu_w=- \frac{K}{\mu_w}\ k_{r,w}\ \vec\nabla P_w\;,
\end{equation}
and
\begin{equation}
\label{eqn8}
\vec \nu_n =- \frac{K}{\mu_n}\ k_{r,n}\ \vec\nabla P_n\;.
\end{equation}
Here $\nu_w$ and $\nu_n$ are the Darcy velocities of the wetting and
non-wetting fluids, $\mu_w$ and $\mu_n$ the viscosities of the wetting
and non-wetting fluids, $K$ is the permeability of the porous
medium. $k_{r,w}(s)$ and $k_{r,n}(s)$ are the relative permeabilities
of the wetting and non-wetting fluids.  $s$ is the non-wetting
saturation.  One distinguishes between the pressure in the wetting
fluid $P_w$ and in the non-wetting fluid $P_n$.  They are related
through the capillary pressure $P_c(s)$ by
\begin{equation}
\label{eqn9}
P_n-P_w=P_c\;.
\end{equation}
The relative permeabilities and the capillary pressure are assumed to
be functions of the non-wetting fluid saturation $s$ alone.  These
equations treat the porous medium as a continuum.  There are three
functions entering these three equations that are determined by the
physics at the pore level: $k_{r,w}(s)$, $k_{r,n}(s)$ and $P_c(s)$.

An alternate recent theory \cite{hsbksv16} based on thermodynamics
\cite{kb08,kbjg10} proposes the relations
\begin{equation}
\label{eqn10}
\frac{d\vec \nu}{ds}=\vec \nu_n-\vec \nu_w\;,
\end{equation}
and 
\begin{equation}
\label{eqn11}
s\ \frac{d\vec \nu_n}{ds}+(1-s)\ \frac{d\vec \nu_w}{ds}=0\;,
\end{equation}
where $\vec \nu= s \vec \nu_n+(1-s)\vec \nu_w$ is the saturation-weighted
average Darcy velocity.  The pore-level physics enters the picture
through the constitutive equation $\vec \nu=\vec \nu(s,\vec\nabla P)$,
where $P$ is the pressure.

The functions $k_{r,w}(s)$, $k_{r,n}(s)$, $P_c(s)$, or in the last
case $\vec \nu=\vec \nu(s,\vec\nabla P)$ are macroscopic functions;
they are defined at the continuum level.  Other theories will have
other macroscopic functions that connect the pore level physics to the
continuum level. Such functions are the results of the collective
behavior of the fluids in vast numbers of pores.  To be able to
calculate the precise behavior of the fluids in a small number of
pores as done when using Lattice Boltzmann method is not enough to
determine fully the physics on large scales.  This is well-known in
other fields such as statistical mechanics where longe-range
correlations that are generated by the short-range interactions
between the microscopic components may dominate the behavior.

It is therefore tempting to develop a {\it statistical mechanics\/}
for immiscible two-phase flow in porous media.  The goal of
statistical mechanics is precisely to bind the microscopic and the
macroscopic worlds together, and it has been very successful doing
this in the past. We will in this paper attempt to take the first
steps in this direction.

In the 1950s through 60s, work was done on a statistical description
of flow in porous media \cite{cc50,s54,s65,a66}.  Since the late 80s,
a theory of two-phase flow using a thermodynamic approach has been
developed and employed by Hassanizadeh and Gray
\cite{hg90,gh98,g99,g89}.  More recently, Valavanides and Daras
\cite{vd16} employ tools from statistical mechanics to describe flow.
Hansen and Ramstad \cite{hr09} proposed to develop a thermodynamical
description of immiscible two-phase flow in porous media based on the
configurations of the fluid interfaces, an approach that is related to
that of Valavanides and Daras.

In the spirit of Hansen and Ramstad \cite{hr09}, we aim to develop a
statistical description of the flow of two immiscible fluids through a
two-dimensional network by constructing a macroscopic description that
applies to the ensemble-averaged behavior of all connected links.
Sinha et al.\ \cite{shbk13} derived a statistical description of
steady state two-phase flow in a single capillary tube. They showed
that the well-known Washburn equation could be derived from the
entropy production in the tube.  They verified that the system was
ergodic and derived an analytical expression for the ensemble
distribution.  The ensemble distribution is the probability
distribution of finding the center of mass of a bubble of the
non-wetting liquid at a particular position in the tube. The ensemble
distribution in the one-dimensional case was found to be inversely
proportional to the velocity of the non-wetting bubble.  Given that a
slow bubble stays proportionally longer in a link, this velocity
dependence is self-evident.  This idea was developed further, by
demonstating in \cite{sshbkv16} that the probability for a given
configuration of interfaces in a network, not just a one-dimensional
one, is proportional to the inverse of the total flow through the
network. This probability distribution was then used to form the basis
for a Markov chain Monte Carlo method for sampling configurations in a
network model.

Whereas the {\it configurational probability distribution\/} that was
derived in Savani et al.\ \cite{sshbkv16}, gave the probability
density for the interfaces between the fluids forming a given
configuration in the {\it entire network,\/} we will here construct an
{\it ensemble distribution\/} for the {\it individual links.\/} That
is, we will derive the joint probability density for any link in the
network to have a given saturation, that the non-wetting fluid it contains
will have its center of mass at a given position and that its radius will
have a given value. 

We consider here for concreteness a network of pores each
characterized by a length and a radius.  We define flow velocity and
saturations for each link and set up the joint statistical
distribution between these and the radius distribution. We will assume
that the porous medium --- the network of pores --- is homogeneous.
This implies that if there are two statistically similar networks, the
combined system will have the same properties as the separate systems.

The paper is organized as follows. In section \ref{sec2} we describe
the porous medium model we will use for the theoretical
derivations. We use a biperiodic square lattice where the links model
the pores. This simplifies the theoretical discussion while retaining the
complexity of the flow.  Section \ref{sec3} introduces the ensemble
distribution that provides the joint probability distribution for pore
radius, pore saturation, and the position of bubbles in the pores.
The first of these variables characterizes the porous medium whereas
the other two characterize the flow.  We go on to demonstrate that the
ensemble distribution is inversely proportional to the volume flow
through the links. We also demonstrate that the system is ergodic.
The next section \ref{ave} connects the ensemble distribution with
dfferent macroscopic quantities, namely the fractional volume flow,
the saturation, the pressure difference and the entropy production.
In section \ref{sec5} we test numerically some of the central results
of the previous sections.  Our conclusions are given in section
\ref{sec7}.

\section{Defining the Variables Characterizing the Flow and the Porous Medium}
\label{sec2}

In the same way as Bakke and {\O}ren \cite{bo97,s11} extracted a
network from the pore space of a porous medium, we replace the
original porous medium by a network representing its pore space. All
our variables will be defined with reference to the links in this
network.

\begin{figure}
\includegraphics[width = .5\textwidth,clip]{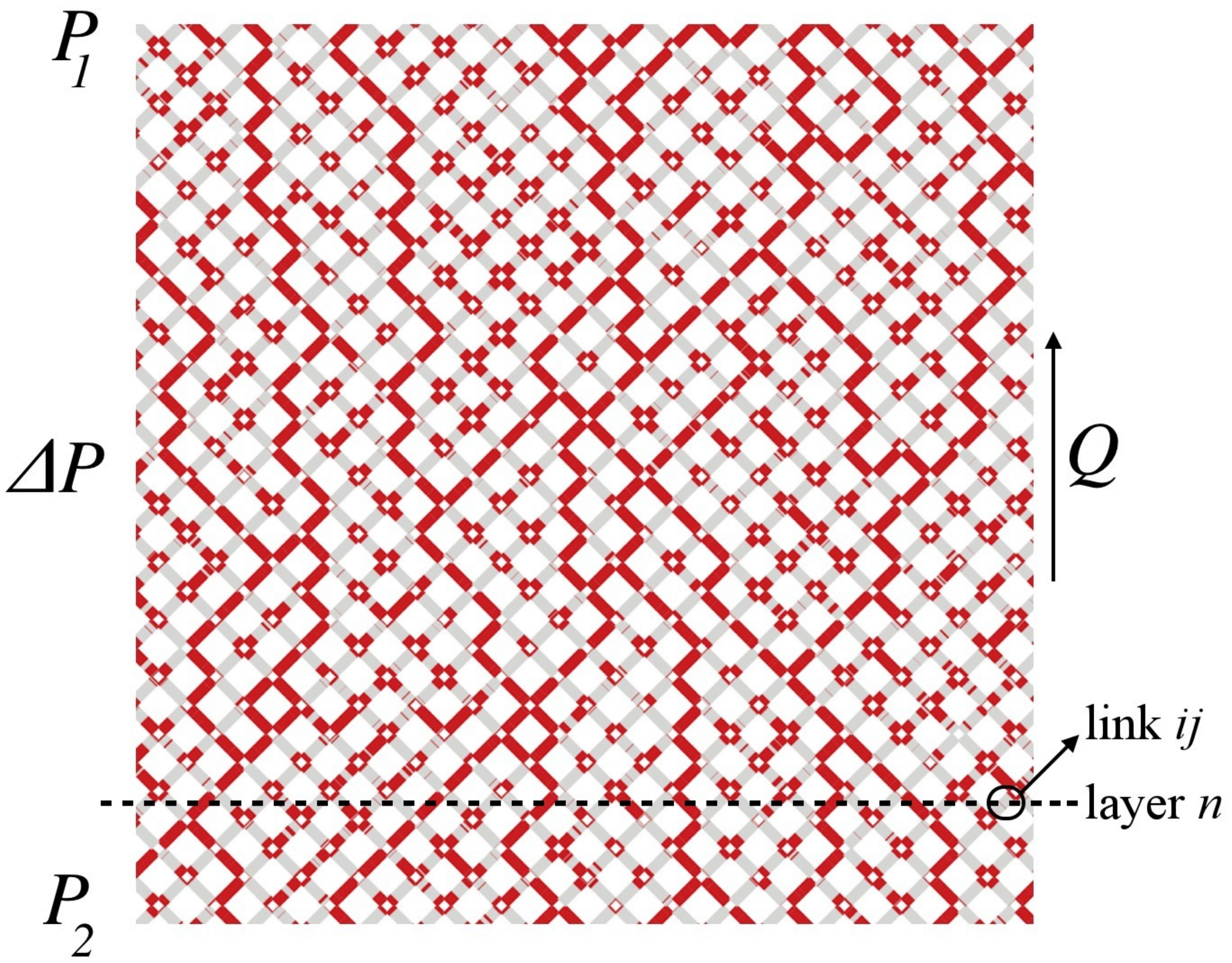}
\caption{The square network with a typical steady-state configuration
  of wetting (white) and non-wetting (red) fluids.  We control the
  volume flow $Q$ through the network.  This generates a pressure
  difference $\Delta P=P_1-P_2$ which depends on the saturation $s$,
  between the top and bottom of the network. Note that the network is
  biperiodic: fluid which escapes at the top enters the corresponding
  node at the bottom and fluid which leaves to the right enters at the
  left.}
\label{fig1}
\end{figure}

In this paper, however, we go one step further and consider a lattice
in the form of a square grid.  This is of course a considerable and
unrealistic simplification compared to the topology of a real pore
network.  However, as the goal of this work is {\it not\/} to consider
a given structure but to develop a {\it general theory,\/} it is
convenient to use the square network.  It simplifies the discussion
while retaining the important subtleties.  The square lattice is
periodic in both directions.  We orient it so that the main axes form
$45^\circ$ with the average flow direction, see Fig.\ \ref{fig1}.
There are $L\times L$ distinct links in the lattice. This means that
there are $L$ distinct layers of links, see Fig.\ \ref{fig1}. For a
square lattice, the total number of nodes is then $L^2/2$ and a link
between two neigboring nodes $i$ and $j$ is denoted by $ij$ where
$i,j\in[0,L^2/2]$.

We assume that all links in the network have the same length $l$.  The radius 
$r_{0,ij}$ varies from link to link and is drawn from a spatially uncorrelated 
distribution $f_r(r_{0,ij})$.  

A volume flow $Q$ across the network in the vertical direction generates
a pressure difference across one layer $\Delta P/L$ depending on $s$ in the 
opposite direction, see Fig.\ \ref{fig1}.

We define the non-wetting saturation $s_{ij}$ in link $ij$ as
\begin{equation}
s_{ij}\equiv \frac{v_{n,{ij}}}{v_{ij}}\;,
\end{equation}
where $v_{n,ij}$ refers to the volume of non-wetting fluid in the link and 
$v_{ij}$ is the total volume of the link.

We assume that the wetting fluid does not wet the pores completely so
that it does not form films.  The non-wetting fluid will form bubbles
that fill the cross-sectional area of the links.  The position of the
bubbles may be characterized by one number since they cannot pass or
change their distance from each other. We characterize their motion
through the time derivative of one position variable $x_{b,ij}$ which
signifies e.g.\ the position of their center of mass measured along
the link of length $l$. Hence $0 \le x_{b,ij} \le l$.

We will consider steady-state flow \cite{esthfm13}.  Experimentally, this
is attained when the two immiscible fluids are injected simultaneously into
the porous medium with all control parameters kept constant and all 
measured macroscopic quantities fluctuate around well-defined and constant
averages.  In our square lattice, the steady state is attained when the 
fluids are allowed to circulate long enough in the biperiodic network.      
The steady state does {\it not\/} imply that the interfaces at the pore
level are static.  Rather, $\Delta P $ may be so high that all interfaces
move and the system would still be in the steady state.

\section{Ensemble Distribution}
\label{sec3}

At a given moment in time, there is a certain configuration of the two
fluids in the network. The ensemble distribution is derived from this
snapshot and is considered to be a time-independent probability
distribution of the two fluids over the ensemble of links since the
flow appears under steady-state conditions.  The exact nature of the
ensemble distribution is still unknown, however, the aim of this work
is to derive some of its properties. Such knowledge will enable the
integration across the network for determining various properties of
interest. For instance, we are interested in the average pressure
difference and the fractional volume flow of the wetting and the
non-wetting fluids when the total volume flow $Q$ and $s$ are imposed.
In particular we are interested in the total volume flows of each of
the single fluids.  Thermodynamics on the macroscopic level has
recently been used to relate these quantities \cite{hsbksv16}.
However, this approach is entirely macroscopic.  Using an ensemble
distribution, we can build a bridge between the properties of a single
link and the overall performance of the network.  We aim to develop a
new method that can solve the up-scaling problem in the context of
immiscible multi-phase flow in porous media.

The ensemble distribution we develop in the following is at the {\it
  indivual link level.\/} That is, pick a link in the network at
random. What is the joint probability density that this link has a
given radius, saturation and the center-of-mass position of the
non-wetting bubbles it contains is at a certain position.

As was stated in the introduction, this is different from the
configurational probability distribution, that is the probability
density for the interfaces between the two fluids takes on a given
configuration in the network, that was derived in Savani et
al.\ \cite{sshbkv16} and used to construct a Markov chain Monte Carlo
algorithm for sampling configurations in network models.

\subsection{The One-Dimensional Distribution}
\label{sec3.1}

Working towards the goal to determine the ensemble distribution beyond
one dimension, we start with conclusions from a one-dimensional
sequence of links \cite{shbk13}.  We have reported earlier that the
probability that a bubble has a certain position $x_{b,ij}$ in the
link, is inversely proportional to the velocity $dx_b/dt$ of the fluid
in that link
\begin{equation}
\Pi (x_b)=\frac{1}{\tau }
\frac{1}{dx_b/dt}=\frac{1}{l}\frac{\langle q\rangle}{%
q(x_b)}\;,\text{ \ \ \ for \ \ }0<x_b<l\;,  
\label{config1d}
\end{equation}
where $\tau =\pi {r_0}^{2}l/\langle q\rangle$\ is the average 
time the bubble takes to move from one end of the link to the other end, 
and $\langle q\rangle$ is the average volume flow. This ensemble 
distribution expresses the sensible fact that the time a bubble spends 
in a link, is inversely proportional to its velocity.

The system characterized by ensemble distribution in equation
(\ref{config1d}) is by construction ergodic \cite{shbk13}. The time
average of a function $g=g(x_b)$ is $\overline{g}$, and we have
\begin{equation}
\label{1dergod}
\overline{g}=\frac{1}{\tau}\ \int_0^\tau dt\ g\left(x_b(t)\right)
=\int_0^l\ dx_b\ \frac{g(x_b)}{\tau dx_b/dt}=\int_0^l dx_b\ \Pi(x_b)\ g(x_b)
=\langle g\rangle\;.
\end{equation}
Hence, the time average of any function equals its ensemble average ---
which is the definition of ergodicity.

\subsection{Ensemble distribution in higher dimesions}
\label{sec3.2}

In higher dimensions, at any instance, the state of a link can be
characterized by the center-of-mass position of the bubbles in it,
$x_{b,ij}$, the saturation $s_{ij}$, and the radius $r_{0,ij}$ of the
link.  In the course of time, a single link will see the passage of
many bubbles with different sizes. One may calculate the time average
of $q_{ij}$ for each individual link.

A subsequent average over the radii of the links returns the average
volume flow $\langle q\rangle$ in the links.  A fast bubble will spend
proportionally less time in a given link than a slow bubble.  This is
true whether the link is part of a one-dimensional or a
multi-dimensional network.  This suggests that also in the
multi-dimensional case, the ensemble distribution,
$\Pi(x_{b,ij},s_{ij},r_{0,ij})$, will be inversely proportional to the
volume flow $\left\vert {
  q_{ij}(x_{b,ij},s_{ij},r_{0,ij})}\right\vert$.  By the same argument
that led to equation (\ref{1dergod}) and hence, ergodicity, the
multidimensional system must be ergodic.

A general form of the ensemble distribution is 
\begin{align}
\Pi (x_{b,ij},s_{ij},r_{0,ij})& 
=\frac{\langle\left\vert q\right\vert\rangle}{%
\left\vert q_{ij}(x_{b,ij},s_{ij},r_{0,ij})\right\vert }%
f(x_{b,ij},s_{ij},r_{0,ij})  \label{configa} \\
& \text{for}\ 0<x_{b,ij}<l\ \text{and}\ 0<s_{ij}<1\;,  \notag
\end{align}
where $f(x_{b,ij},s_{ij},r_{0,ij})$ is assumed to be normalized. The
distribution $f(x_{b,ij},s_{ij},r_{0,ij})$ may, in principle, depend
on the volume flow $q_{ij}(x_{b,ij},s_{ij},r_{0,ij})$.  If
$f(x_{b,ij},s_{ij},r_{0,ij})$ does not depend on $x_{b,ij}$\ via its
flow dependence, the implication is that
$f(x_{b,ij},s_{ij},r_{0,ij})$$=f(s_{ij},r_{0,ij})/l$.  It then follows
that the ensemble distribution takes the form
\begin{align}
\Pi (x_{b,ij},s_{ij},r_{0,ij})& 
=\frac{1}{l}\frac{\langle\left\vert q\right\vert\rangle%
}{\left\vert q_{ij}(x_{b,ij},s_{ij},r_{0,ij})\right\vert }f(s_{ij},r_{0,ij})
\nonumber \\
& \text{for}\ 0<x_{b,ij}<l\ \text{and}\ 0<s_{ij}<1\;.  \label{config}
\end{align}%
The function $f(s_{ij},r_{0,ij})$, which is also normalized, 
gives the joint distribution of the saturation and the link radii.

\section{From ensemble distribution to macroscopic quantities}
\label{ave}

The aim of this section is to calculate the fractional volume flow of 
the non-wetting fluid, the saturation, the pressure drop
and the entropy production, all macroscopic variables, from the
ensemble distribution.

\subsection{Average Absolute Volume and Fractional Volume Flows}
\label{sec4.1}

The average absolute volume flow is given by 
\begin{equation}
 \langle\left\vert q\right\vert \rangle\ =\int_{0}^{\infty
}dr_{0,ij}\int_{0}^{1}ds_{ij}\int_{0}^{l}dx_{b,ij}\
\Pi (x_{b,ij},s_{ij},r_{0,ij})\left\vert
q_{ij}(x_{b,ij},s_{ij},r_{0,ij})\right\vert\;.
\end{equation}
The form can be verified by introducing the general ensemble
distribution in equation (\ref{configa}), and using that $f$ is
normalized.  It turns out that the form of the general ensemble
distribution in equation (\ref{configa}) is sufficient to obtain this
result.

The total absolute volume flow in the direction of the pressure
difference through one layer (see Fig.\ \ref{fig1}) is equal to
\begin{equation}
\left\vert Q'\right\vert \equiv \sum_{ij}\ \left\vert
q_{ij}(x_{b,ij},s_{ij},r_{0,ij})\right\vert\;,
\label{9}
\end{equation}
where the summation is over all the links $ij$ in that particular
layer.

The total absolute volume flow through the cross section or through
each layer is the same for incompressible fluids. Hence $\left\vert
Q'\right\vert = \left\vert Q\right\vert$. The average is equal to
\begin{equation}
\langle\left\vert Q\right\vert 
\rangle=L\langle\left\vert q\right\vert \rangle\;,  
\label{18}
\end{equation}
where $L$ is the number of links in a layer. The last equality expresses the
fact that the links in a layer form an ensemble of links with the ensemble
distribution given in equation (\ref{config}.) 

We proceed to calculate the absolute fractional flow through the
system. The average of the total absolute volume flow of the
non-wetting fluid in the direction of the overall pressure difference
is equal to,
\begin{equation}
\begin{split}
\langle\left\vert Q_{n}\right\vert \rangle& \equiv L\langle\left\vert
q_{n,ij}(x_{b,ij},s_{ij},r_{0,ij})\right\vert \rangle \\
& =L\langle s_{ij}\left\vert q_{ij}(x_{b,ij},s_{ij},r_{0,ij})\right\vert 
\rangle\;.
\end{split}%
\end{equation}
With the help of the ensemble distribution, the absolute flow of the
non-wetting fluid equals
\begin{equation}
\begin{split}
\langle\left\vert Q_{n}\right\vert \rangle& =L\int_{0}^{\infty
}dr_{0,ij}\int_{0}^{1}ds_{ij}\int_{0}^{l}dx_{b,ij} \\
& \times \Pi (x_{b,ij},s_{ij},r_{0,ij})s_{ij}\left\vert
q_{ij}(x_{b,ij},s_{ij},r_{0,ij})\right\vert \\
& =L\langle\left\vert q\right\vert \rangle\int_{0}^{\infty
}dr_{0,ij}\int_{0}^{1}ds_{ij}\int_{0}^{l}dx_{b,ij} \\
& \times s_{ij}f(x_{b,ij},s_{ij},r_{0,ij}) \\
& =L\langle\left\vert q\right\vert \rangle\langle s\rangle\;.
\end{split}
\label{avqnw}
\end{equation}

The non-wetting fraction of the absolute volume flow is then equal to
\begin{equation}
F \equiv \frac{\langle\left\vert Q_{n}\right\vert \rangle}{\langle\left\vert
Q\right\vert \rangle}
= \int_{0}^{\infty
}dr_{0,ij}\int_{0}^{1}ds_{ij}%
\int_{0}^{l}dx_{b,ij}s_{ij}f(x_{b,ij},s_{ij},r_{0,ij})=\langle s\rangle\;.  
\label{21}
\end{equation}
The fraction of the total absolute non-wetting volume flow, is
therefore equal to the ensemble average of the degree of
saturation. Again, the form of the general ensemble distribution
equation (\ref{configa}) is sufficient to obtain this result. The
relation can be tested numerically and experimentally with information
of the distributions. We show that it is obeyed for a particular
network in the end of the paper.

Equation (\ref{21}) is at a first glance surprising.  However, it should be
remembered that the average saturation is {\it per link\/} and not per
volume.

\subsection{Average Saturation}
\label{avSat}

The volume average of the saturation of the links in any layer is given by%
\begin{equation}
\begin{split}
s& \equiv \frac{V_{n}}{V}=\frac{\sum_{j}v_{ij}s_{ij}}{\sum_{j}v_{ij}} \\
& =\frac{\int_{0}^{\infty
}dr_{0,ij}\int_{0}^{1}ds_{ij}\int_{0}^{l}dx_{b,ij}v_{ij}s_{ij}\Pi
(x_{b,ij},s_{ij},r_{0,ij})}{\int_{0}^{\infty
}dr_{0,ij}\int_{0}^{1}ds_{ij}\int_{0}^{l}dx_{b,ij}v_{ij}\Pi
(x_{b,ij},s_{ij},r_{0,ij})} \\
& =\frac{\langle vs\rangle}{\langle v\rangle}\;.
\label{eqn17}
\end{split}%
\end{equation}%
In the one-dimensional sequence of links, all the links have the same 
volume $v_{ij}=\pi lr_{0,ij}^{2}$, so that $\langle s\rangle=s$. This 
implies that the fraction of the total absolute non-wetting volume flow 
is given by $F=s$. This is generally not the case in multi-dimensional 
systems, except when there are no capillary forces.

An interesting observation is that when the distribution of the 
saturation and the radius are not correlated, it follows that
\begin{equation}
f(s_{ij},r_{0,ij})\equiv
\int_{0}^{l}dx_{b,ij}f(x_{b,ij},s_{ij},r_{0,ij})
=f_{s}(s_{ij})f_{r}(r_{0,ij})\;.
\label{corrEq}
\end{equation}
This implies that 
\begin{equation}
\label{eqn19}
F=\langle s\rangle=s\;.
\end{equation}
At high capillary numbers, one has that $F=s$ since the capillary forces
play no role.  This implies equation (\ref{eqn19}) is valid in the
high-capillary number regime.

\subsection{Average Pressure Difference}
\label{sec4.3}

In experiments or simulations in which the volume flow is controlled,
the pressure difference cannot be fixed.  The fluctuating driving force 
follows from equation \eqref{wash}
\begin{equation}
\begin{split}
& \Delta p_{ij}(x_{b,ij},s_{ij},r_{0,ij})-p_{c,ij}(x_{b,ij},s_{ij},r_{0,ij}) \\
& =-\frac{8\mu _{av,ij}(s_{ij})}{\pi r_{0,ij}^{4}}%
q_{ij}(x_{b,ij},s_{ij},r_{0,ij})\;,
\label{eqn100}
\end{split}%
\end{equation}
where $p_{c,ij}(x_{b,ij},s_{ij},r_{0,ij})$ is the capillary pressure drop due to
interfaces in the link and $\Delta p(x_{b,ij},s_{ij},r_{0,ij})$ is the
pressure drop across the link and 
$\mu_{av,ij}=s_{ij}\mu_n+(1-s_{ij})\mu_{w}$ is the volume-weighted average
viscosity.  Using the ensemble distribution, the absolute average driving 
force is given by 
\begin{equation}
\begin{split}
& \left\langle\left\vert \Delta p-p_{c,ij}\right\vert \right\rangle 
= \frac{8}{\pi }\int_{0}^{\infty
}dr_{0,ij}\int_{0}^{1}ds_{ij}\int_{0}^{l}dx_{b,ij} \\
& \times \Pi (x_{b,ij},s_{ij},r_{0,ij})\frac{\mu _{av,ij}(s_{ij})}{%
r_{0,ij}^{4}}\left\vert q_{ij}(x_{b,ij},s_{ij},r_{0,ij})\right\vert\;.
\end{split}%
\end{equation}%
By introducing equation (\ref{configa}) for the ensemble distribution, we 
obtain 
\begin{equation}
\begin{split}
\langle\left\vert \Delta p-p_{c,ij}\right\vert \rangle& 
= \frac{8}{\pi }\langle\left\vert
q\right\vert \rangle\int_{0}^{1}ds_{ij}\int_{0}^{\infty }dr_{0,ij} \\
& \times \frac{\mu_{av,ij}(s_{ij})}{r_{0,ij}^{4}}%
f(s_{ij},r_{0,ij})\;.
\end{split}%
\end{equation}%

Using this expression, one can find the overall mobility $\mathrm{M}$  of the 
fluids in the network
\begin{equation}
\mathrm{M} = \frac{\langle\left\vert q\right\vert \rangle}{\langle\left\vert 
\Delta p-p_{c,ij}\right\vert\rangle} \;,
\end{equation}%
which corresponds to Darcys law for the system.
\subsection{The Entropy Production}
\label{sec4.4}

In non-equilibrium thermodynamics, the second law is formulated in terms of 
the entropy production in the system \cite{kb08,kbjg10}.  The entropy 
production quantifies the energy dissipated in the form of heat in the 
surroundings.  In the present case, this amounts to the viscous dissipation. 
According to the second law, the dissipation is always positive. 
The expression for the entropy production in terms of the ensemble 
distribution must obey this condition at a local level, i.e.\ at the 
scale of a single link. For the whole system, we can find the 
average entropy production using equation \eqref{config}, 
\begin{equation}
\begin{split}
T& \left\langle\frac{dS_{\text{irr}}}{dt}\right\rangle=-\int_{0}^{\infty
}dr_{0,ij}\int_{0}^{1}ds_{ij}\int_{0}^{l}dx_{b,ij} \\
& \times \Pi (x_{b,ij},s_{ij},r_{0,ij})q_{ij}(x_{b,ij},s_{ij},r_{0,ij}) \\
&  \times  \left (\Delta p_{ij}-p_{c,ij}(x_{b,ij},s_{ij},r_{0,ij})\right )
\\
& = \langle\left\vert q\right\vert \rangle\int_{0}^{\infty
}dr_{0,ij}\int_{0}^{1}ds_{ij}\int_{0}^{l}dx_{b,ij} \\
& \times f(x_{b,ij},s_{ij},r_{0,ij})\left\vert \Delta
p_{ij}-p_{c,ij}(x_{b,ij},s_{ij},r_{0,ij})\right\vert \\
& = \langle\left\vert q \vert\right\rangle  
\left \langle\vert \Delta p-p_{c,ij}\right\vert \rangle\;.\\
\end{split}%
\label{entropyprod}
\end{equation}%
In the second equality we used that ${ q}_{{\large ij}}${\large \ and
}$(\Delta p-p_{c,ij})$ have the opposite signs in accordance with the
second law of thermodynamics.  We see that the local as well as the
global entropy production have the correct bilinear form. This
confirms that the ensemble distribution given in equation
(\ref{configa}) is the correct choice.

\section{Numerical Verification}
\label{sec5}

We test and develop numerically some of the main results of the
previous sections using the network model described in the Appendix.

The network was initialized with a random configuration of bubbles for
a desired saturation $s$. Measurements were started only after the
system had reached steady state.

We used a spatially uncorrelated uniform distribution on the interval 
$[0.1,0.4]$ mm for the radii.  The length of the links was 1 mm.  The 
non-wetting and wetting model fluids were given the same viscosity 
$\mu $ = 0.1 Pa s. The surface tension $\gamma$ between the fluid was 
set to $30$ mN/m. 

The simulations were performed for two different volume flows that
were kept constant throughout the simulations, $Q=26\text{mm}^3$/s and
$Q=260\text{mm}^3$/s. They corresponded to a capillary number, defined
as
\begin{equation}
\text{Ca}\equiv \frac{Q\mu }{A\gamma }\;,  
\label{Qdef}
\end{equation}
where $A$ is defined as the cross-section of the network given by
$\sum_{ij} \pi r_{0,ij}^2$ where the sum runs over a layer.  
The capillary numbers were Ca = 0.01
and 0.1. The system size was $L\times L=40\times 40$ except in
Figs.\ \ref{fig4}, \ref{fig5} and \ref{fig6} where also $L\times
L=20\times 20$ were used.  Results are averaged over $10$ samples for
each series of measurements with different Ca.

\begin{figure}
\includegraphics[width = .4\textwidth,clip]{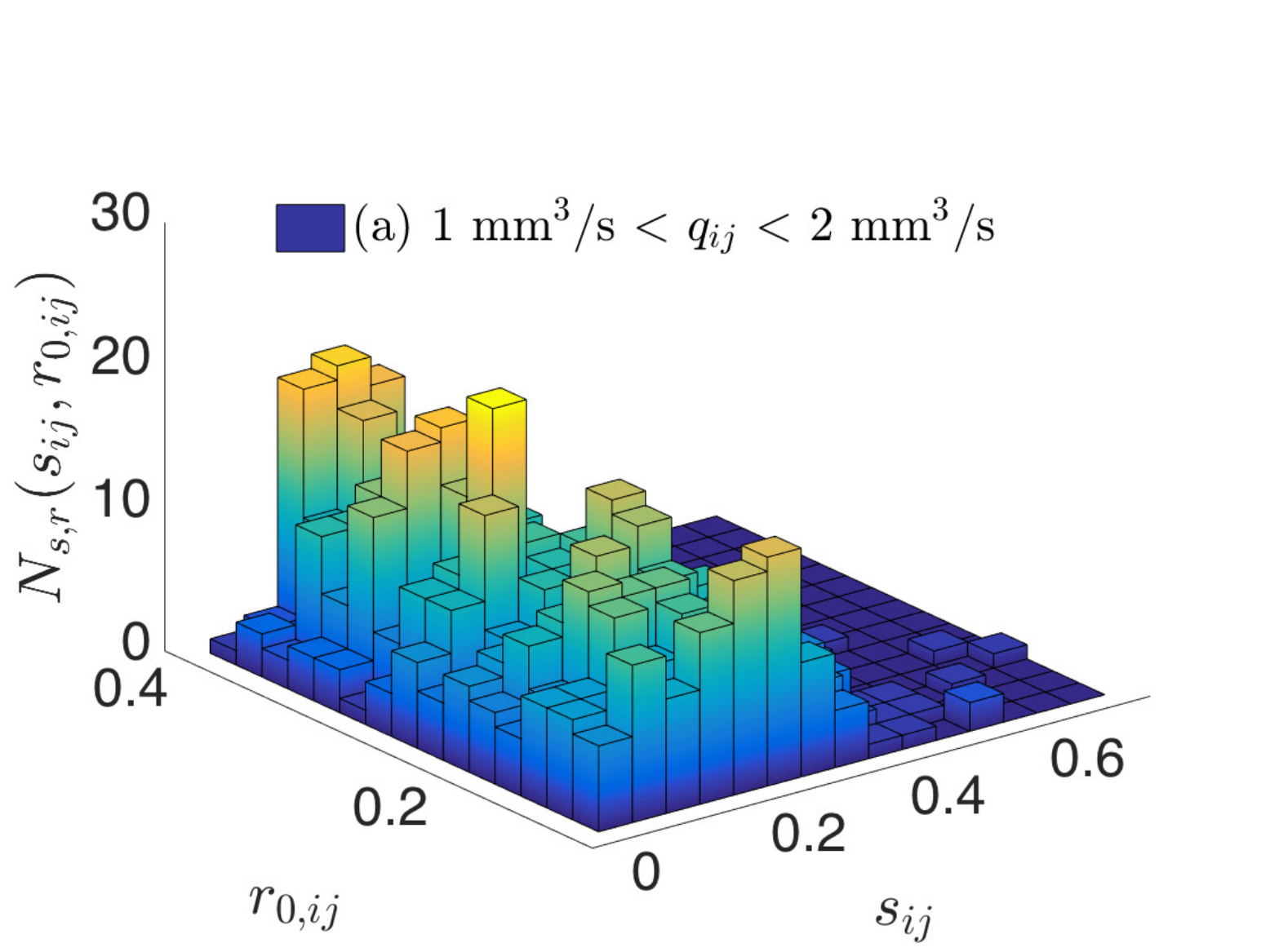}
\includegraphics[width = .4\textwidth,clip]{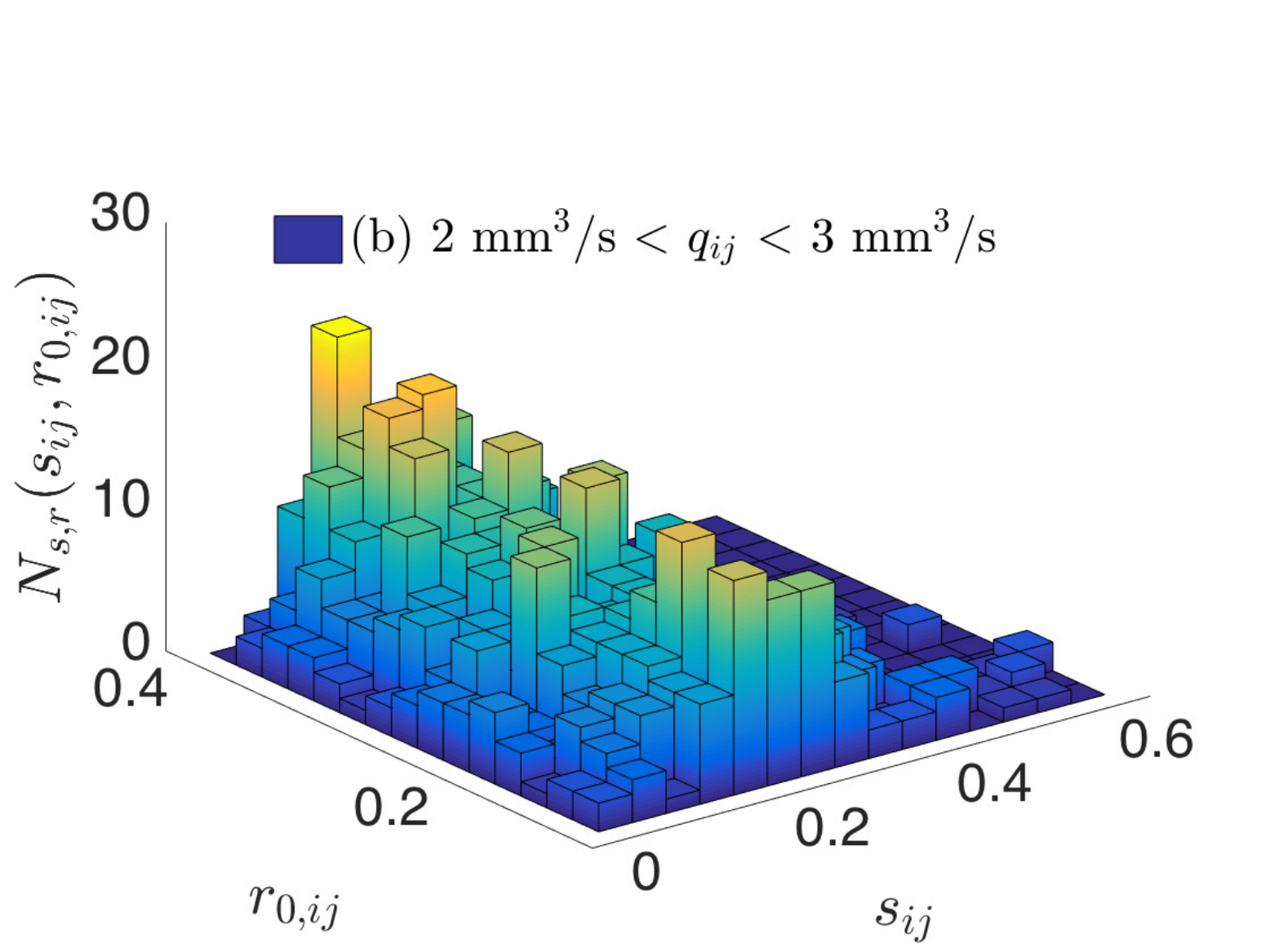}
\includegraphics[width = .4\textwidth,clip]{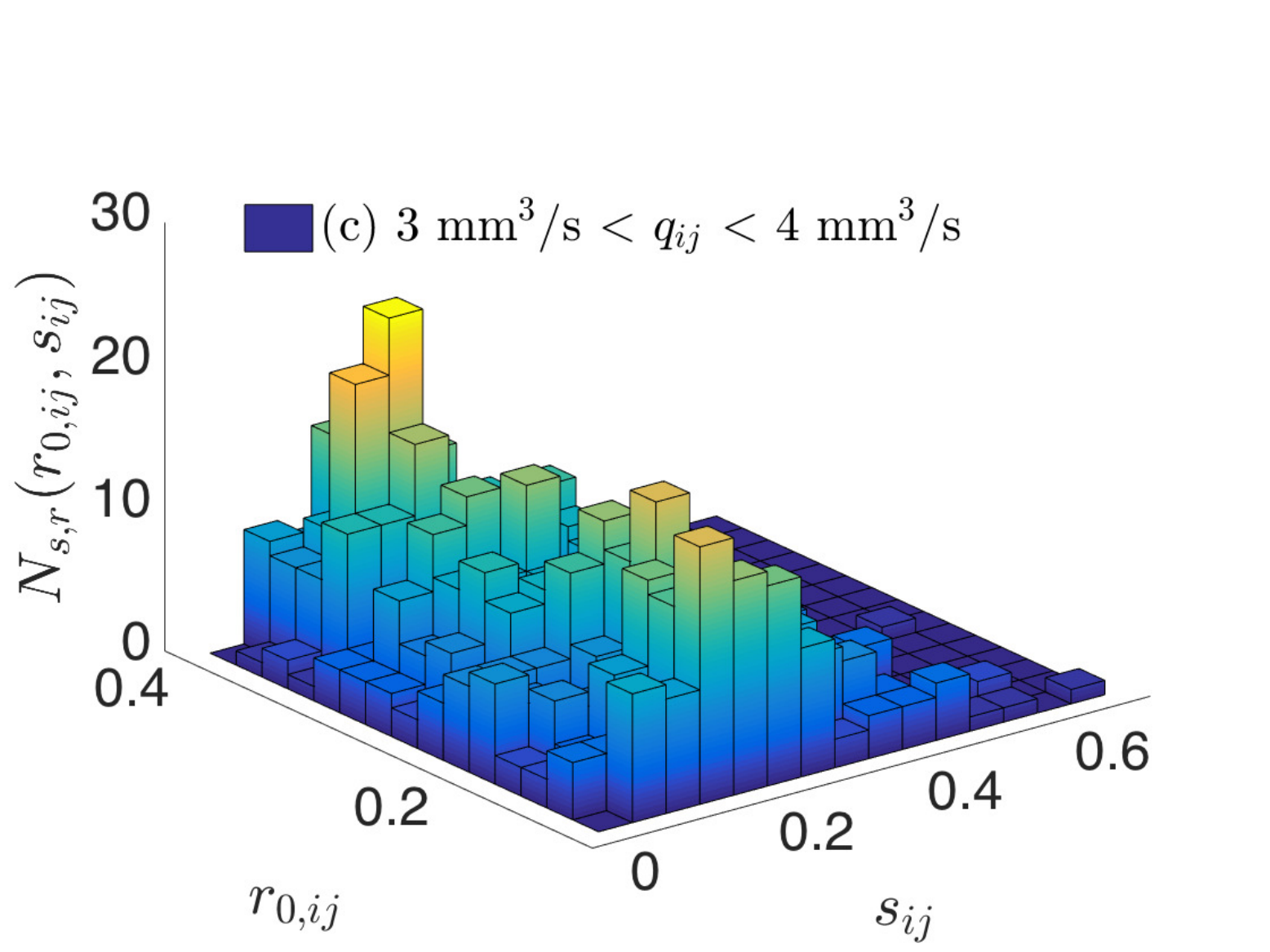}
\includegraphics[width = .4\textwidth,clip]{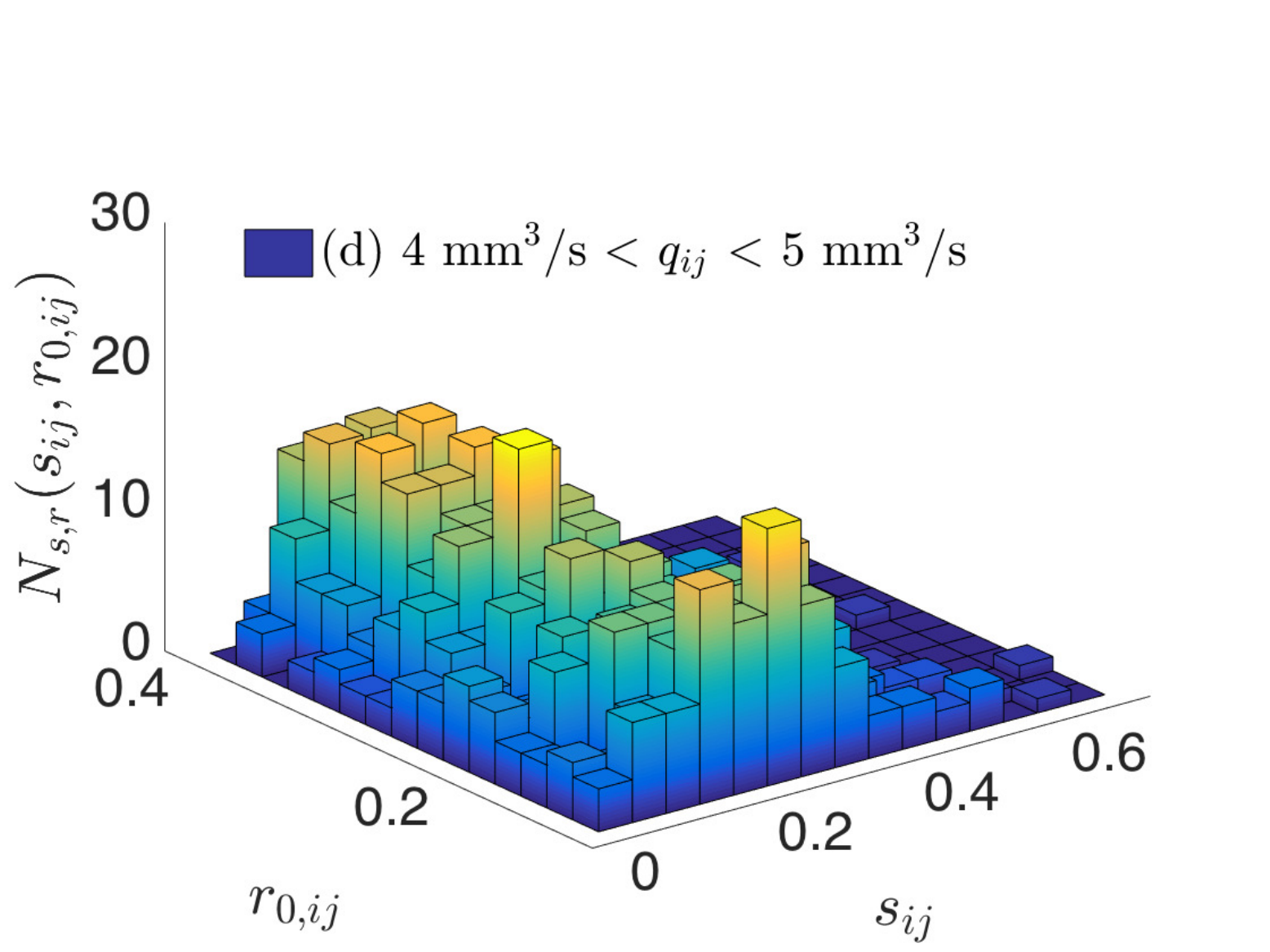}
\caption{Joint histogram $N_{s,r}$ for $r_{0,ij}$ and $s_{ij}$ for the
  given ranges of $q_{ij}$ for Ca = 0.01 and $s=0.2$.}
\label{fig2}
\end{figure}

In equation (\ref{configa}) we give the general form of the
configurational probability.  In Fig.\ \ref{fig2}, we
show histograms $N_{s,r}$ proportional to the joint probability
distribution for $s_{ij}$ and $r_{0,ij}$ when only those links fall
within a narrow range of volume flows are counted. That is, we record
only those links for which (a) 1 mm$^3$/s $< q_{ij}<$ 2 mm$^3$/s, (b)2
mm$^3$/s $< q_{ij}<$ 3 mm$^3$/s, (c) 3 mm$^3$/s $< q_{ij}<$ 4
mm$^3$/s, and (d) 4 mm$^3$/s $< q_{ij}<$ 5 mm$^3$/s. The volume flows
$q_{ij}$ ranged roughly between -2.5 mm$^3$/s and 7.5 mm$^3$/s.  If
equation (\ref{config}) were true, i.e.,
$f(x_{b,ij},s_{ij},r_{0,ij})=f(s_{ij},r_{0,ij})$, then the histograms
in the four figures should be identical.  We see that even though the
features are similar, they are not. Hence, there is an $x_{b,ij}$
dependence in $f(x_{b,ij},s_{ij},r_{0,ij})$ for Ca = 0.01.

\begin{figure}
\includegraphics[width = .6\textwidth,clip]{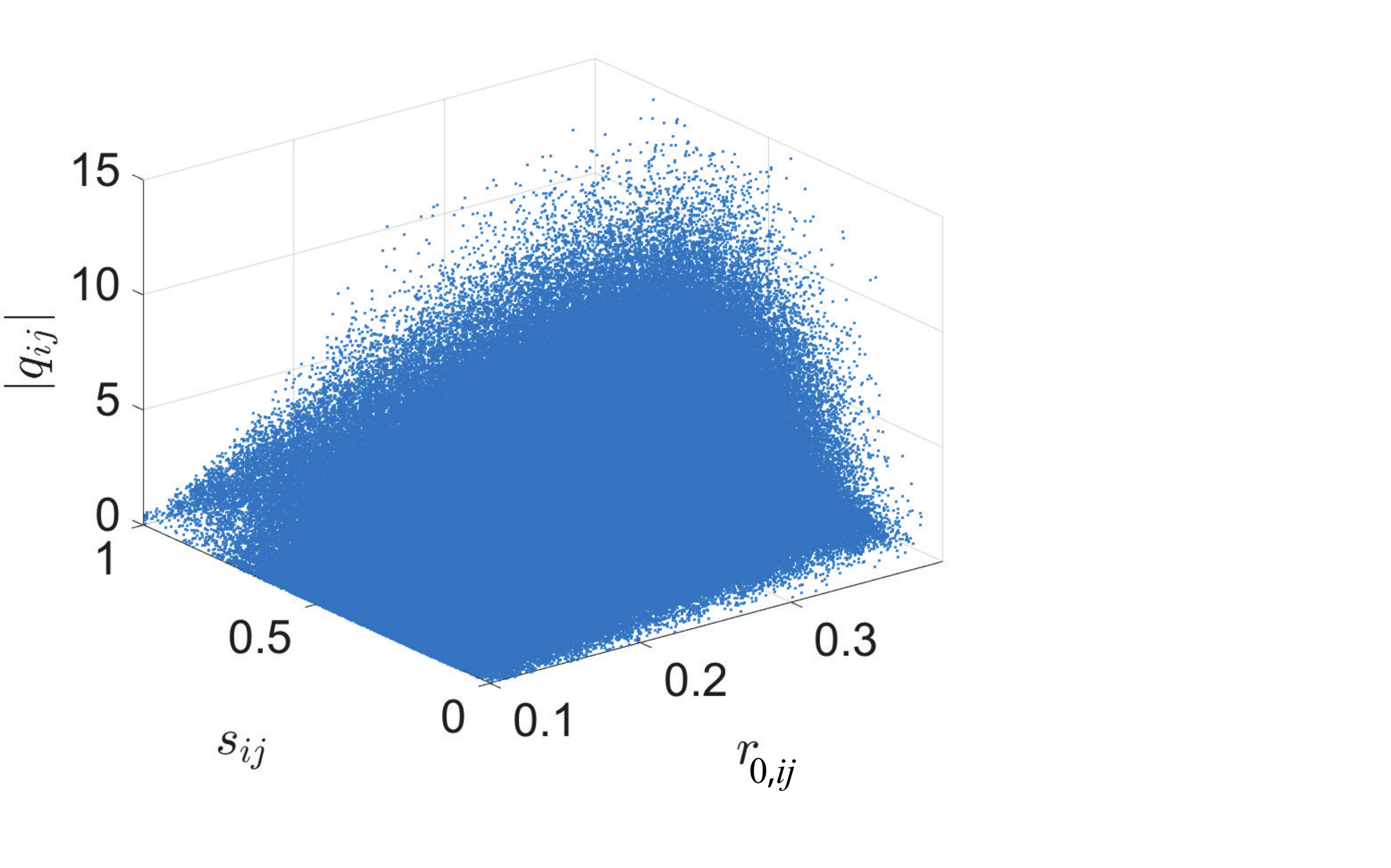}
\caption{A cloud plot of $|q_{ij}|$, $s_{ij}$ and $r_{0,ij}$ for
Ca = 0.01 and $s=0.5$.} 
\label{fig3}
\end{figure}

We may transform the distribution in equation (\ref{configa}) from a
distribution in $x_{b,ij}$ to a distribution in $|q_{ij}|$,
\begin{equation}
\Pi_q (|q_{ij}|,s_{ij},r_{0,ij}) 
=\frac{\langle\left\vert q\right\vert\rangle}{%
\left\vert q_{ij}\right\vert }
f(x_{b,ij}(|q_{ij}|,s_{ij},r_{0,ij}),s_{ij},r_{0,ij})\ 
\left[\frac{\partial}{\partial |q_{ij}|}\right]
x_{b,ij}(|q_{ij}|,s_{ij},r_{0,ij})\;.   
\label{configb}
\end{equation}
We show in Fig.\ \ref{fig3} the cloud of values measured in the system
for Ca = 0.01 and $s=0.5$.  It is this cloud that equation
(\ref{configb}) describes.

\begin{figure}
\includegraphics[width = .4\textwidth,clip]{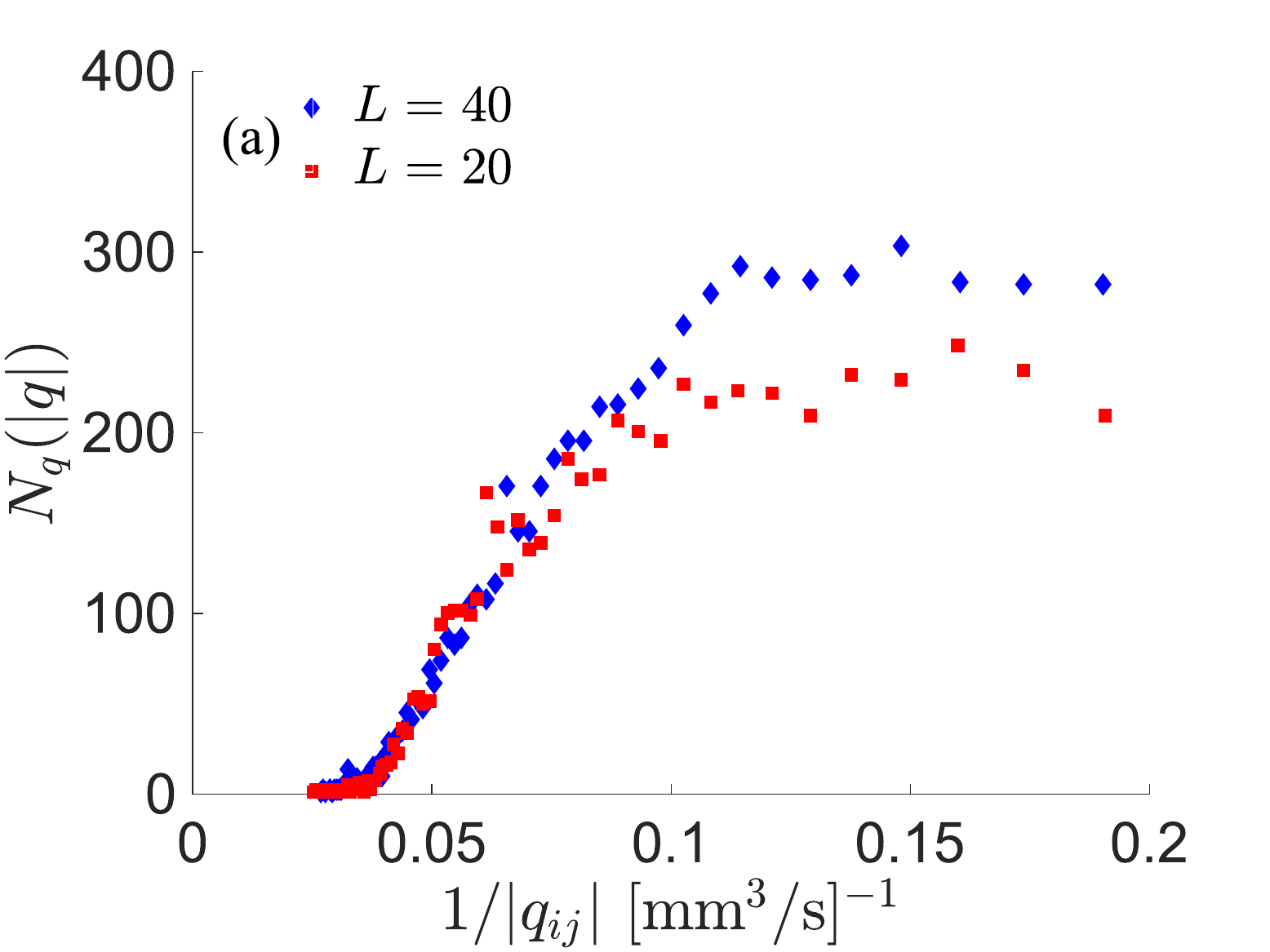}
\includegraphics[width = .4\textwidth,clip]{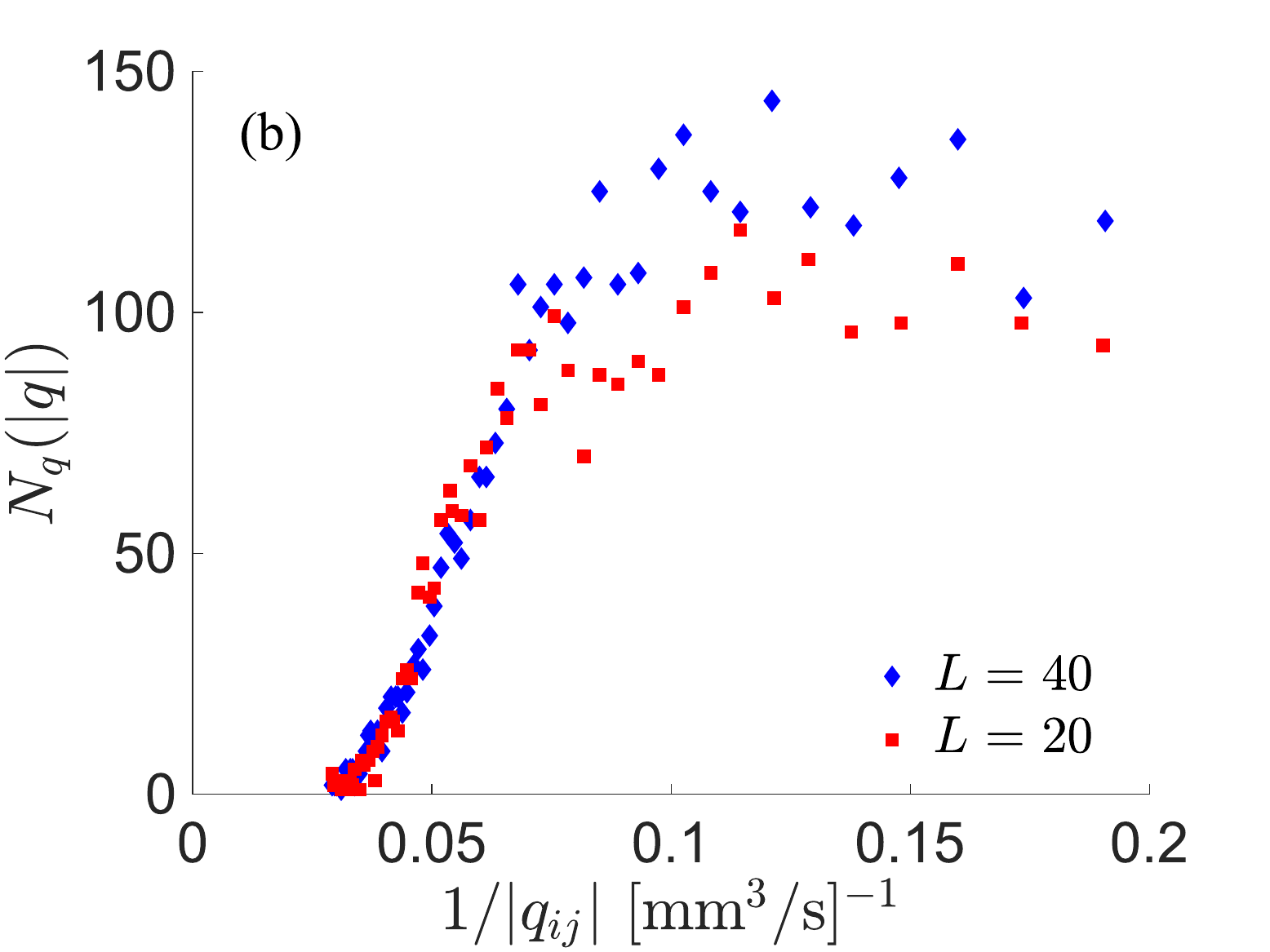}
\caption{Histogram $N_q(|q|)$ of volume flow in links when $0.4 <
  s_{ij} < 0.5$ and $0.2$ mm $< r_{0,ij} < 0.3$ mm for $Ca=0.1$ and
  (a) $s=0.3$ or (b) $s=0.4$.}
\label{fig4}
\end{figure}

We show in Fig.\ \ref{fig4} histograms $N_q$ proportional to $\Pi_q$
for Ca = 0.1 and for two different system sizes, $L\times L = 20\times
20$ and $L\times L=40\times 40$. The volume flows $Q$ have been
adjusted so that the capillary numbers are the same for the two system
sizes.  Only those links with the values of the other two parameters,
$s_{ij}$ and $r_{0,ij}$ within a truncated range have been recorded.
In Figs.\ \ref{fig5} and \ref{fig6} we show the corresponding
histograms for Ca = 0.01.  The histograms for Ca = 0.1 show a gap for
small values of $1/|q_{ij}|$, and for increasing values of
$1/|q_{ij}|$ a somewhat linear region before an essentially flat
region occurs. It is not possible from the results for the two system
sizes to infer a clear trend that could make it possible to
extrapolate the result to infinite system size.  The corresponding
histograms for the Ca = 0.01 case are in Figs.\ \ref{fig5} and
\ref{fig6}.  They are qualitatively different from the histograms for
Ca = 0.1, Fig.\ \ref{fig4}. There is still a gap for small values of
$1/|q_{ij}|$, but from a smallest value, $1/|q_{0,ij}|=1/|\max_{ij}
q_{ij}|$, the histogram raises linearly.  We also note that the $L=40$
data gives a straighter line that the $L=20$ data.  Since the volume
flow $Q$ is kept fixed, there is a largest possible link volume flow
in the system: $\max_{ij} |q_{ij}| = |Q|$, which would occur if $Q$ in
its entirety passed through one link, a possibility that would be more
and more likely the smaller the capillary number due to capillary
blocking.  Hence, for Ca = 0.01, $\Pi_q(|q_{ij}|)$ takes the form
\begin{equation}
\label{eqn1000}
\Pi_q(|q_{ij}|)=
g(s_{ij},r_{0,ij})\left[\frac{1}{|q_{ij}|}-\frac{1}{|Q|}\right]\;,
\end{equation}
where 
\begin{equation}
\label{eqn1001}
g(s_{ij},r_{0,ij})=\langle |q|\rangle
f(x_{b,ij}(|q_{ij}|,s_{ij},r_{0,ij}),s_{ij},r_{0,ij})\ 
\left[\frac{\partial}{\partial |q_{ij}|}\right]
x_{b,ij}(|q_{ij}|,s_{ij},r_{0,ij})\;.   
\end{equation}

\begin{figure}
\includegraphics[width = .5\textwidth,clip]{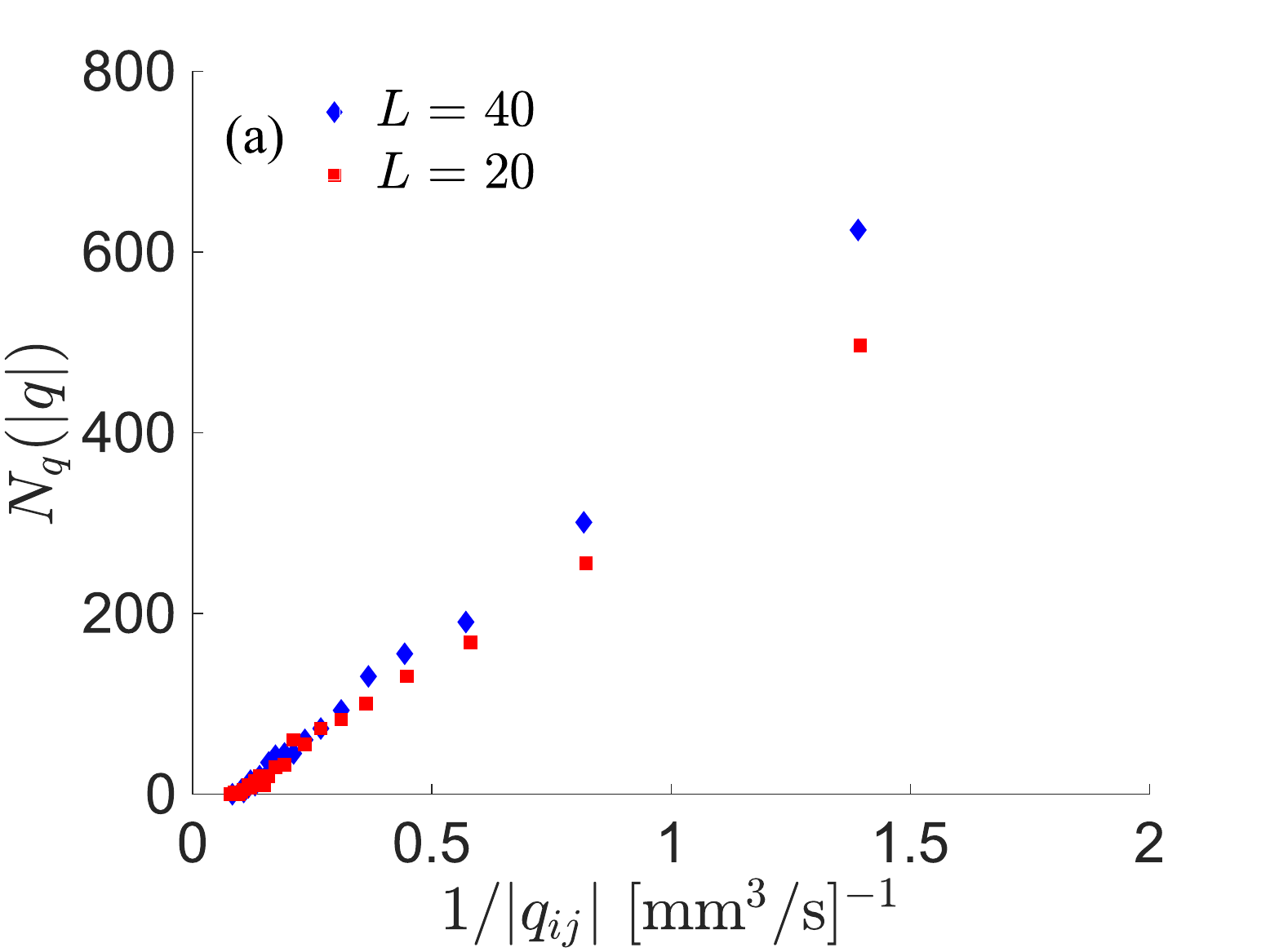}
\caption{Histogram $N_q(|q|)$ of volume flow in links when $0.3 <
  s_{ij} < 0.4$ and $0.2$ mm $< r_{0,ij} < 0.3$ mm for $Ca=0.01$ and
  $s=0.3$.}
\label{fig5}
\end{figure}

The $|Q|$ dependence in equation (\ref{eqn1000}) comes from the use of
the constant-$Q$ ensemble.  If each run had been done with $\Delta P$,
the pressure drop across the network, kept constant, the $1/|Q|$ term
may have vanished.  It was shown in Batrouni et al.\ \cite{bhn87} that
the choice of ensemble; constant-$Q$ or constant-$\Delta P$ had a
profound influence on the high-current end of the current histogram in
the random resistor network, a system that shares some similarity to
the present one.

It should be noted that the immiscible two-phase flow problem
undergoes a phase transition when the saturation is tuned
\cite{rho09}.  For the square lattice, they found the critical
saturation to be $s_c=a+b\log_{10} \rm Ca$, where $a=0.8$ and
$b=0.063$.  For Ca = 0.01, this places the critical point at around
$s_c \approx 0.67$.  In analogy with the random resistor network at
the percolation threshold, we expect $\Pi_q(|q_{ij}|)$ to have a much
more complex form than suggested in equation (\ref{eqn1001})
\cite{arc85,bhl96}, namely that of a multifractal. This has recently
been suggested in connection with immiscible two-phase counterflow in
porous media \cite{zjgz14}.

\begin{figure}
\includegraphics[width = .5\textwidth,clip]{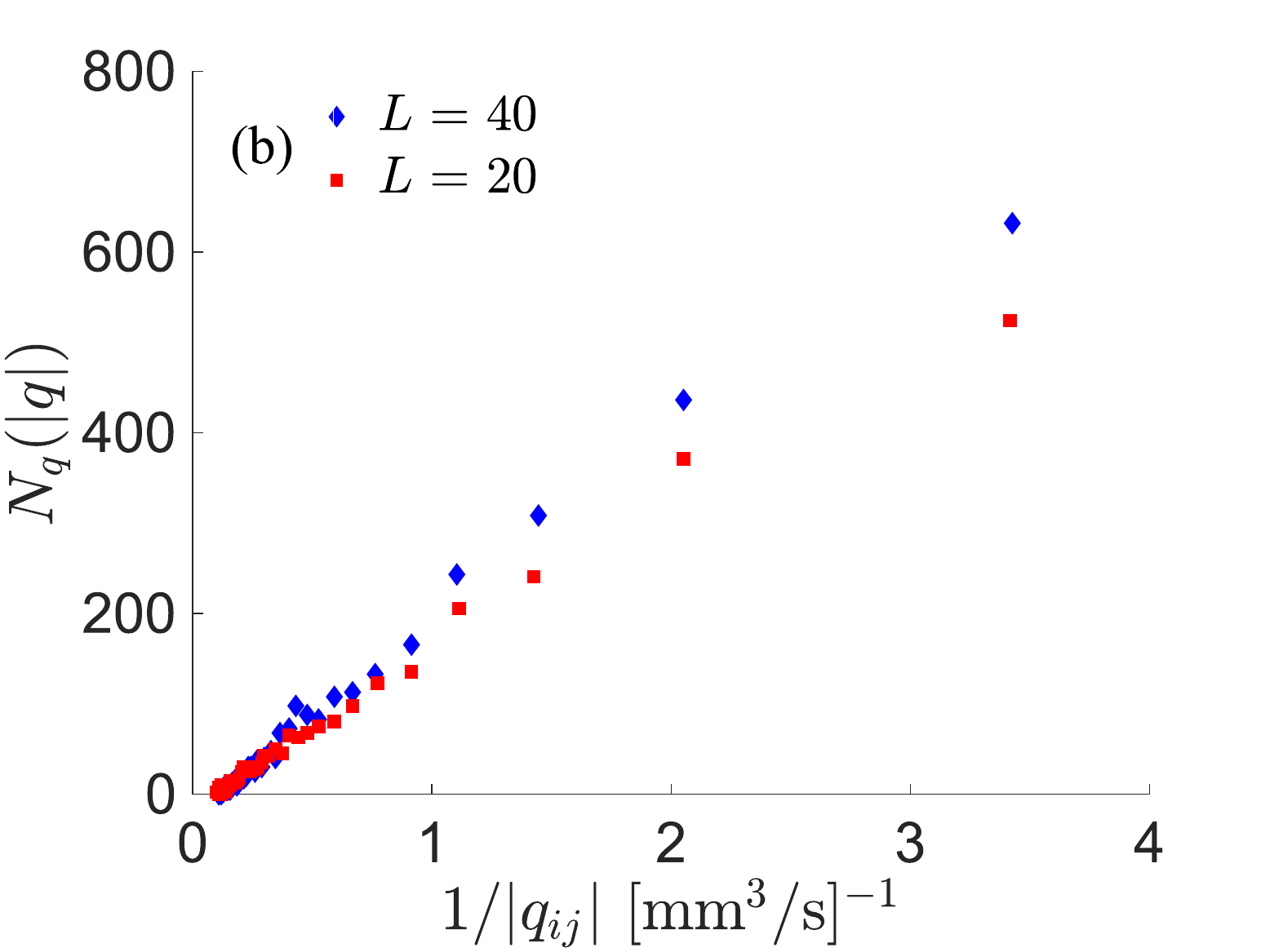}
\caption{Histogram $N_q(|q|)$ of volume flow in links when 
$0.4 < s_{ij} < 0.5$ and $0.2$ mm $< r_{0,ij} < 0.3$ mm for $Ca=0.01$ and 
$s=0.5$.}
\label{fig6}
\end{figure}

\begin{figure}
\includegraphics[width = .45\textwidth,clip]{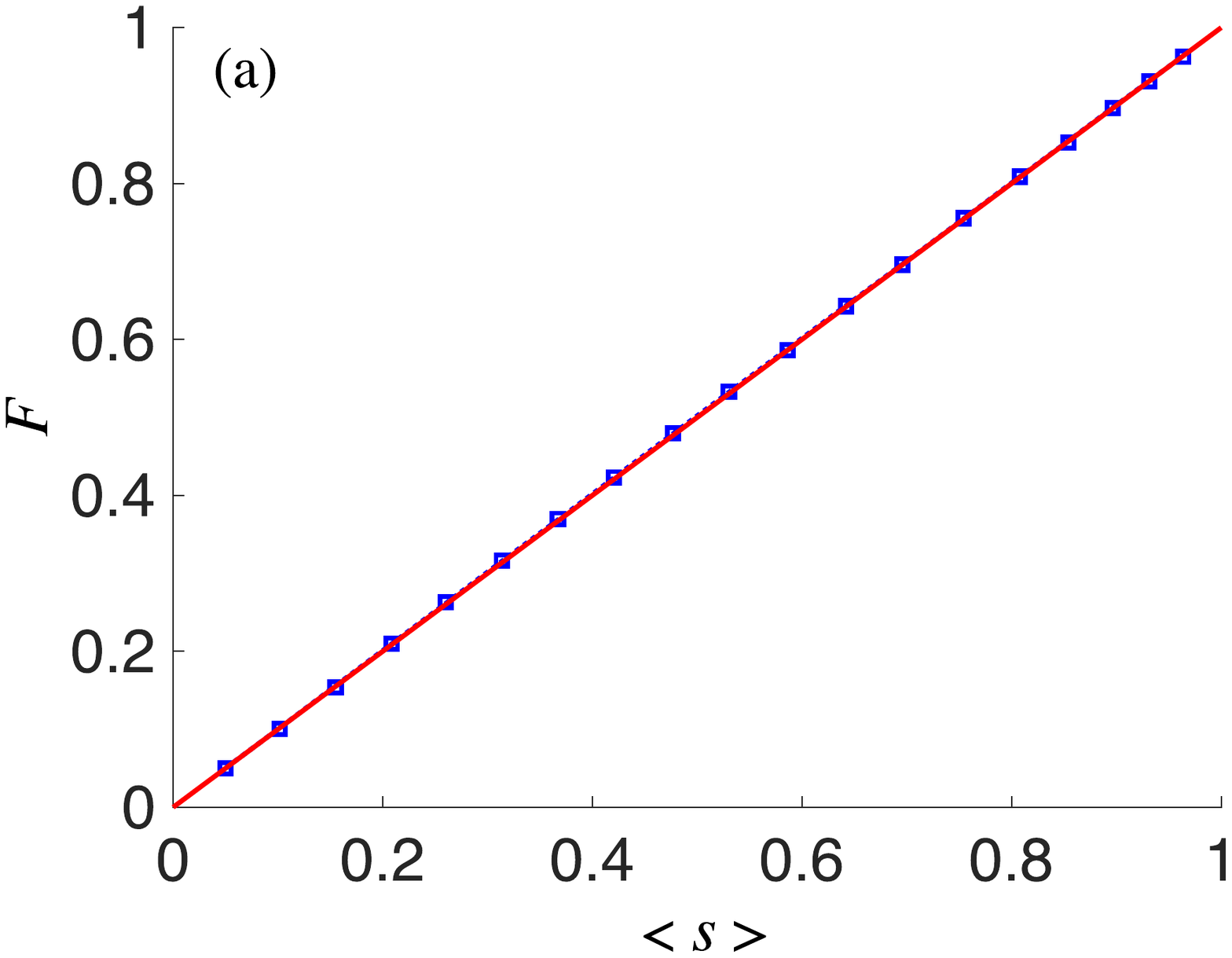}
\includegraphics[width = .45\textwidth,clip]{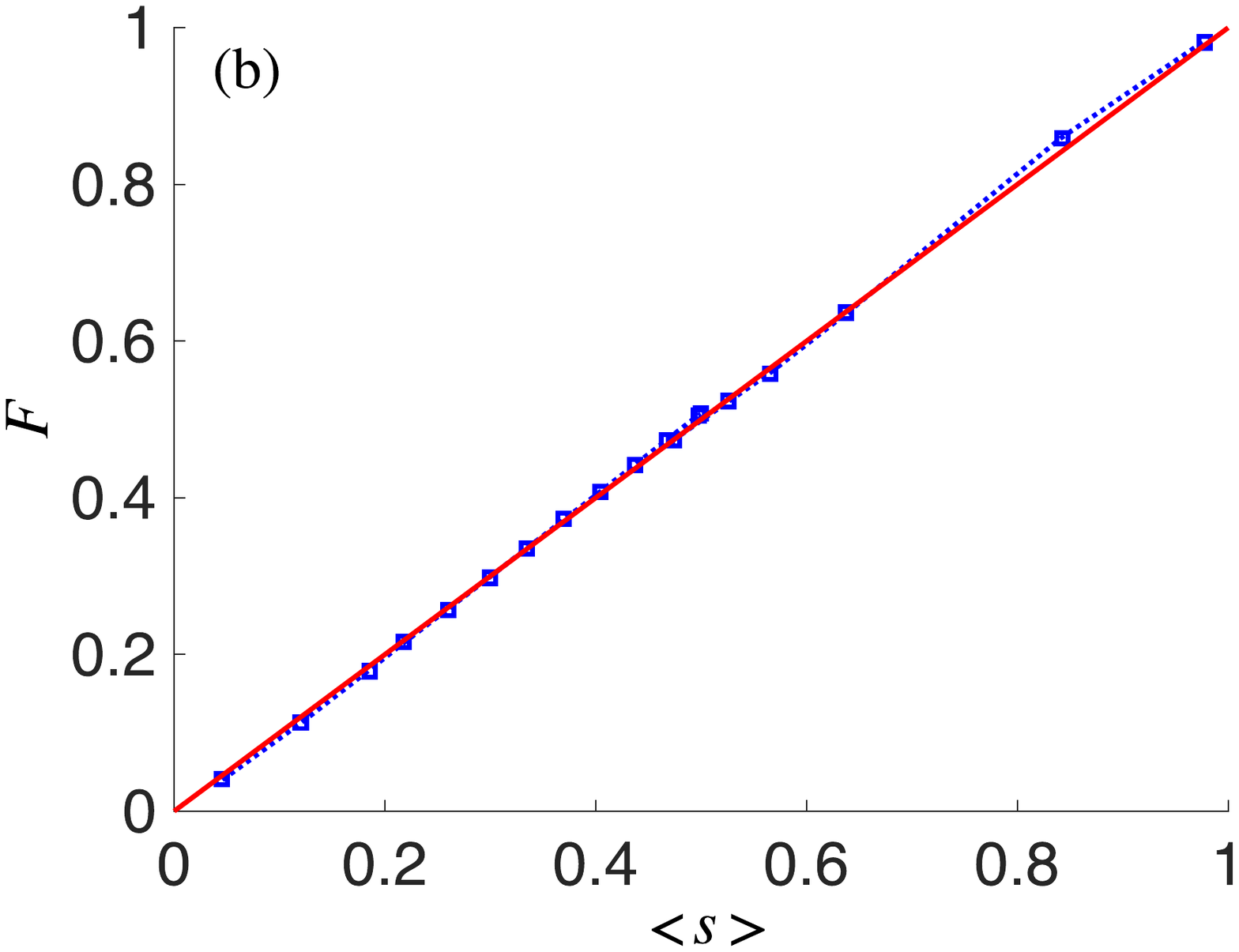}
\caption{The non-wetting fractional flow defined in equation (\ref{21}) $F$
vs.\ the saturation averaged over the links, $\langle s\rangle$
for (a) Ca = 0.1 and (b) Ca = 0.01.  According
to this equation, we expect $F=\langle s\rangle$.  This is also what we 
observe.}
\label{fig7}
\end{figure}

\begin{figure}
\includegraphics[width = .4\textwidth,clip]{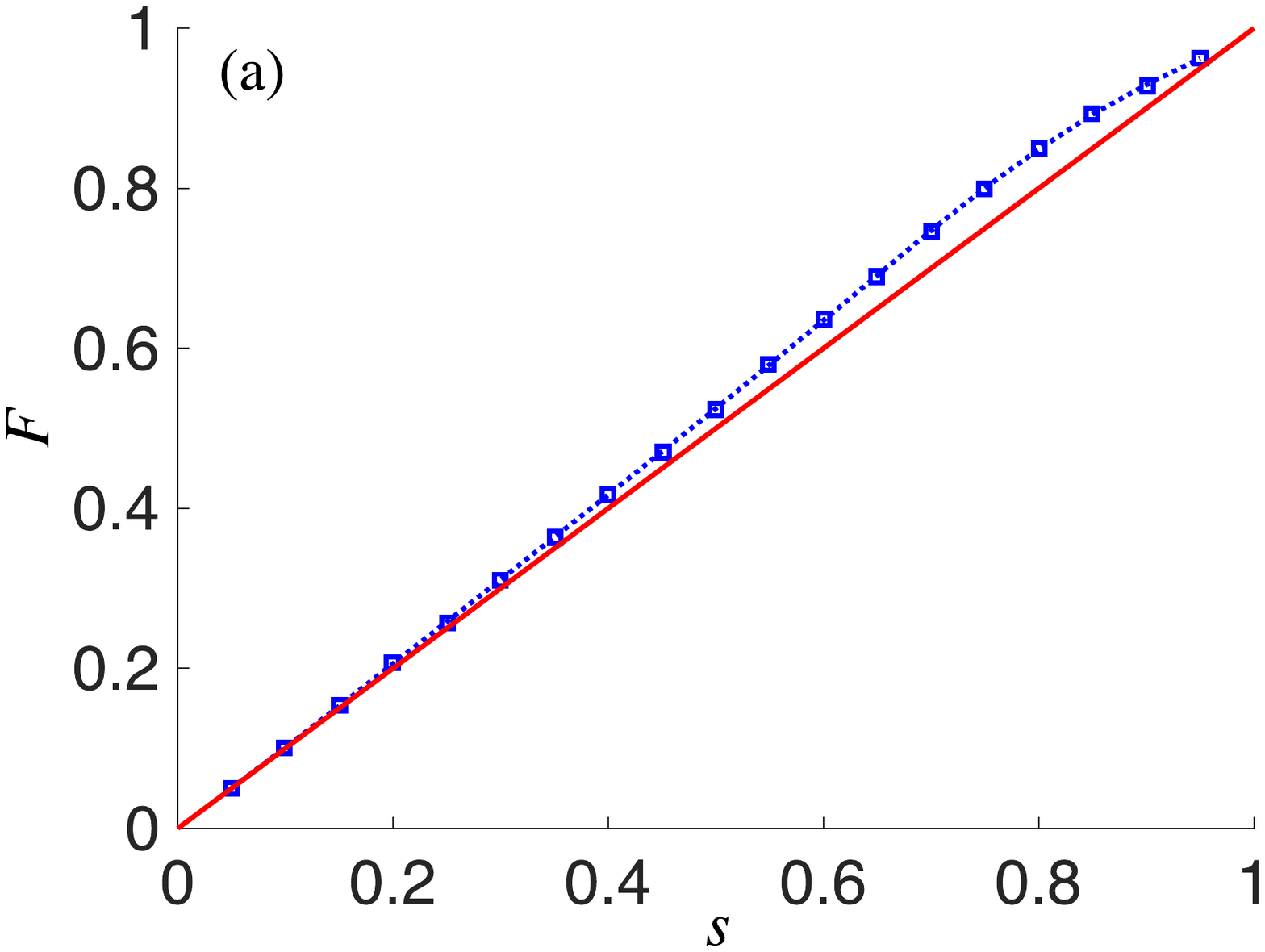}
\includegraphics[width = .4\textwidth,clip]{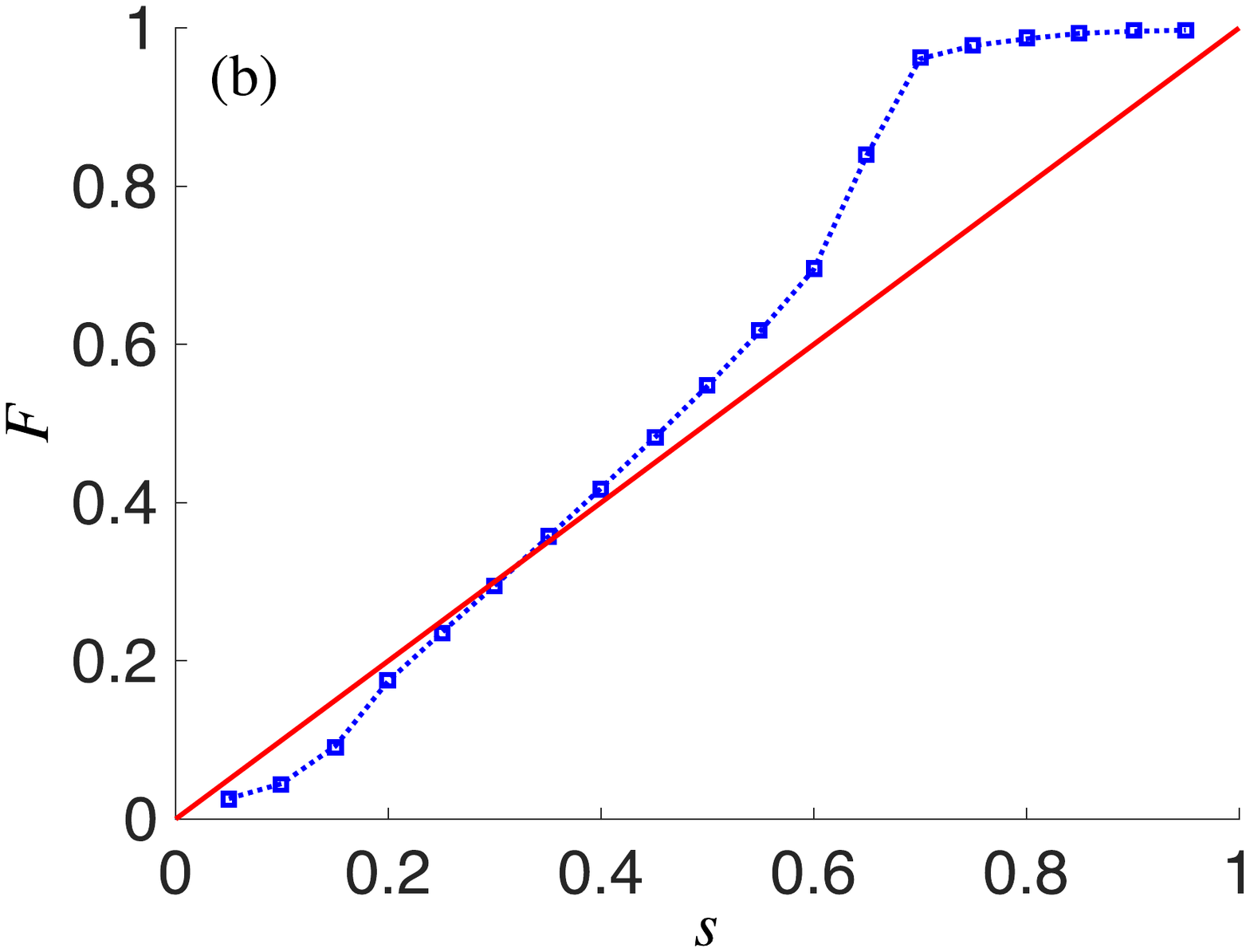}
\caption{The non-wetting fractional flow defined in equation (\ref{21})
$F$ vs.\ the volume averaged saturation $s$ defined in equation (\ref{eqn17})
for (a) Ca = 0.1 and (b) Ca = 0.01.}
\label{fig8}
\end{figure}

In section \ref{ave} it was shown that $F=\langle s\rangle$, equation
(\ref{21}), even if capillary forces are important. We demonstrate the
validity of this calculation in Fig.\ \ref{fig7}.  In Fig.\ \ref{fig8}
we show $F$ as a function of volume-weighted average of the saturation $s$. 
As expected, we see that $F$ is a non-trivial function of $s$.  However,
for larger capillary numbers, $F$ is closer to the diagonal compared to
smaller capillary numbers: compare Fig.\ \ref{fig8}a with \ref{fig8}b.

\begin{figure}
\includegraphics[width = .4\textwidth,clip]{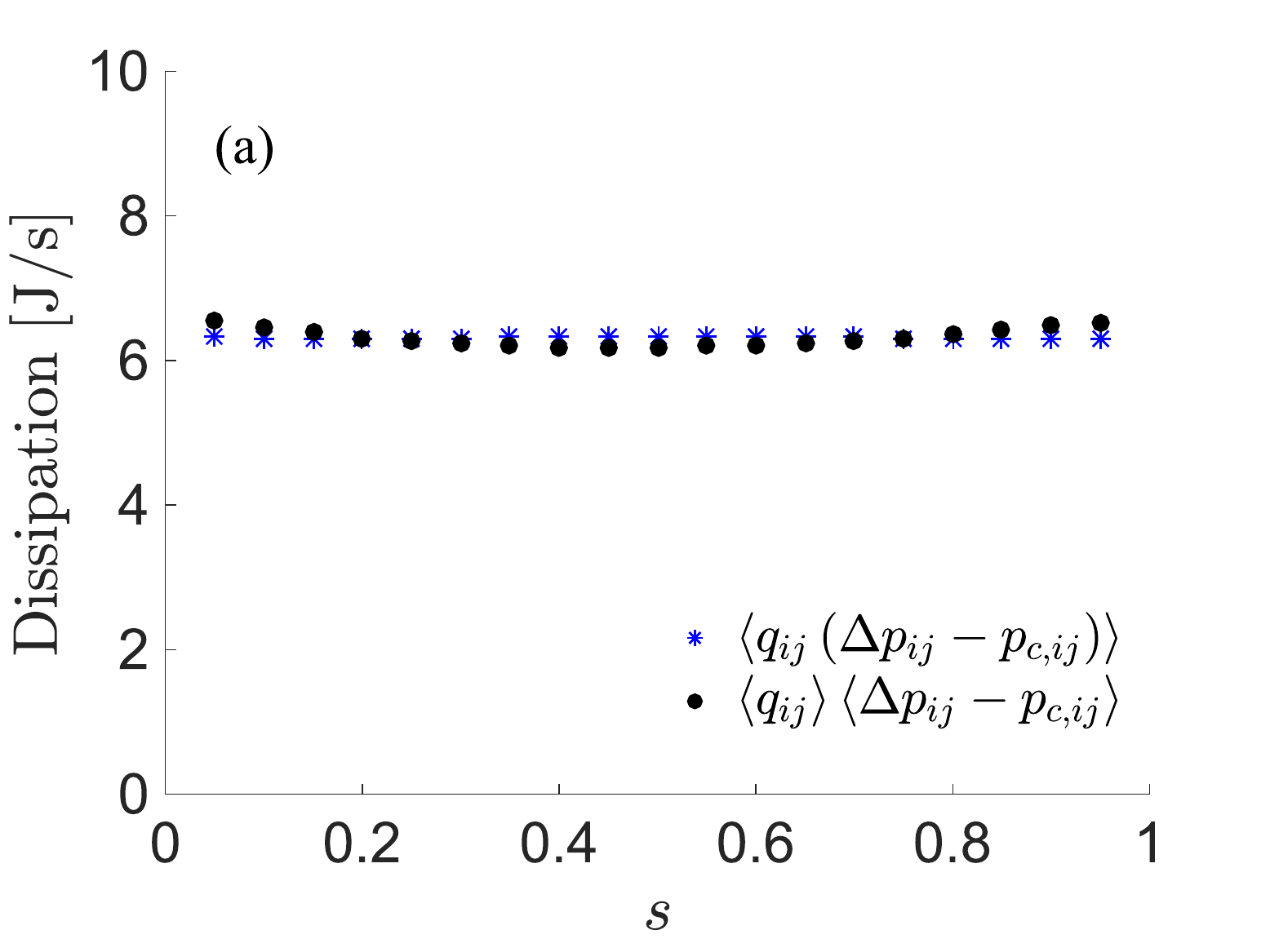}
\includegraphics[width = .4\textwidth,clip]{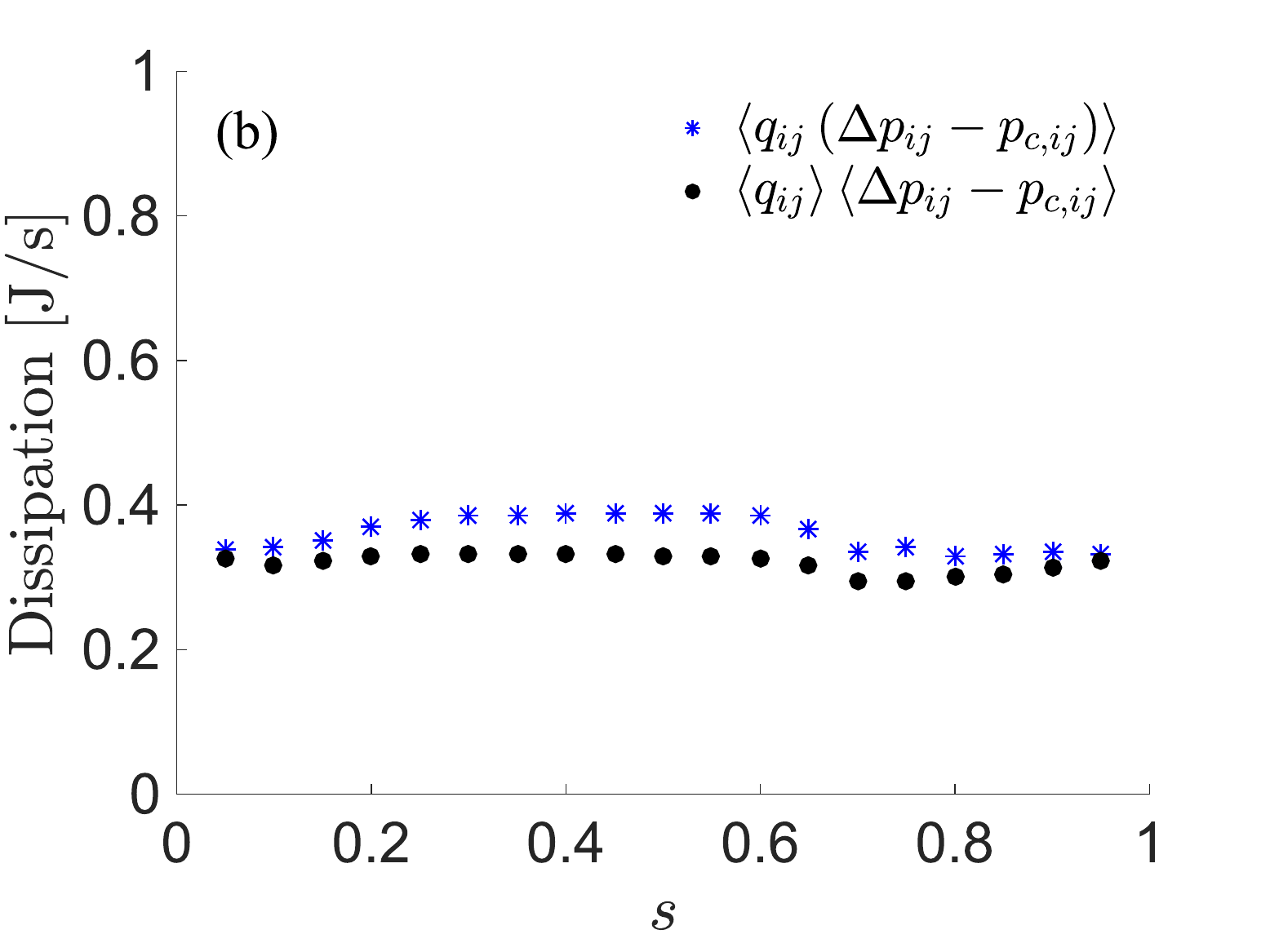}
\caption{The dissipation $\langle q_{ij}(\Delta p_{ij}-p_{c,ij})\rangle$ and
$\langle |q|\rangle \langle |\Delta p -p_{c,ij}|\rangle$ plotted against the
non-wetting saturation for (a) Ca =0.1 and (b) Ca = 0.01.  Equation
(\ref{entropyprod}) predicts that the two data sets should be equal.}
\label{fig9}
\end{figure}

Lastly, we check equation (\ref{entropyprod}) in section \ref{sec4.4}  
in Fig.\ \ref{fig9}.  That is, we plot 
$\langle q_{ij}(\Delta p_{ij}-p_{c,ij})\rangle$ and $\langle |q_{ij}|\rangle
\langle |(\Delta p_{ij}-p_{c,ij})|\rangle$ as a function of the saturation
$s$.  The prediction  of equation (\ref{entropyprod}) is that the two 
quantities should be the same.  For Ca=0.1 (Fig.\ \ref{fig9}a), this 
works well. For the smaller capillary number Ca=0.01 (Fig.\ \ref{fig9}b),
they match to within some 15\% or better.

\section{Conclusion and Perspective}
\label{sec7}

We have presented a statistical mechanical analysis of immiscible
two-phase flow in porous media through the introduction and analysis
of the ensemble distribution $\Pi(x_{b,ij},s_{ij},r_{0,ij})$ which
gives the joint probability density between center-of-mass position of
the bubbles, the saturation and the radius of any link in the network
that models the porous medium.  With the ensemble distribution, any
quantity that does not require the relative positions of the links to
be taken into account may be calculated.  We have presented a few
examples in section \ref{ave}, among them the fractional flow, the
average saturation, the average pressure difference leading to the
effective mobility, and lastly the entropy production or dissipation.
Questions that cannot be answered within this approach are e.g., the
relative statistical weight of a given configuration of interfaces
within the system which e.g.\ comes up in connection with the
construction of the Markov chain Monte Carlo for sampling fluid
configurations. For such questions, the configurational probability
density is necessary\cite{sshbkv16}.

The ensemble distribution $\Pi(x_{b,ij},s_{ij}, r_{0,ij})$ introduced
here relates the dynamics of the system to a probability distribution
that does not contain time.  This is possible due to the ergodicity of
the system, see section \ref{sec3}.  However, since the probability
distribution does not contain time, it cannot answer questions that
have to do with time explicity, e.g., questions concerning correlation
time is outside the realm of this approach.

The ensemble distribution may be transformed into the link volume flow
distribution $\Pi_q(|q_{ij}|,s_{ij},r_{0,ij})$.  This has a
surprisingly simple form for moderate capillary numbers, see equation
(\ref{eqn1000}).  This probability distribution is a close relative of
the current distribution function in the random resistor network that
was extensively studied and shown to be multifractal in the eighties
\cite{arc85,bhl96}.  We expect to find similar complications in the
present system when the saturation is at the critical point studied by
Ramstad et al. \cite{rho09}. 

\begin{acknowledgements}
IS, SK and AH thank VISTA, a collaboration between Statoil and the Norwegian 
Academy of Science and Letters, for financial support. MV thanks the NTNU for 
financial support.  SS thanks the Norwegian Research Council, NFR, and the 
Beijing Computational Science Research Center, CSRC, for financial support.  
The numerical calculations were made possible through a grant of computer time 
by NOTUR, the Norwegian Metacenter for Computational Science.
\end{acknowledgements}

\appendix
\section{Network Model}
\label{seca}

We use the network model shown in Fig.\ \ref{fig1} \cite{amhb98,kah02} to
test some of the central ideas presented in this work. The
links represent cylindrical tubes of varying average radii, containing the 
volume of both the pores and the pore throats of the porous medium.  Each
link is hour-glass shaped so that the capillary pressure due to an interface
at position $0 \le x \le l$ in the link given by 
\begin{equation}
|p_{c,ij}(x)|=\frac{2\gamma\cos\theta}{r_0}\left[1-\cos\left(\frac{2\pi x}{l}\right)
\right]\;,    
\end{equation}
where $\gamma$ is the surface tension.  The volume flow in each link is 
related to the pressure difference across it by the Washburn equation
\begin{equation}
q_{ij}=-\frac{\pi r_{0,ij}^{4}}{8\mu _{av,ij}}(\Delta
p_{ij}-p_{c,ij})\;,  \label{wash}
\end{equation}
$\Delta p_{ij}$ is the pressure difference 
across the link and $p_{c,ij}$ is the sum of the capillary pressure 
contribution from all the interfaces in the link. For a link with a single 
bubble with center-of-mass position $x_{b,ij}$ and saturation $s_{ij}$, and 
surface tension $\gamma$, the capillary pressure on the bubble is given by 
\cite{shbk13}
\begin{equation}
p_{c,ij}(x_{b,ij},s_{ij},r_{0,ij})=\frac{4\gamma}{r_{0,ij}}\sin (\pi
s_{ij})\sin (\frac{2\pi x_{b,ij}}{l})\;, 
\label{eq5}
\end{equation}

At each node volume flow is conserved. This implies that the sum of the 
contributions from equation \eqref{wash} are 
conserved at the nodes. This results in a matrix equation 
for the pressure field. After solving this equation we can use equation 
\eqref{wash} to calculate the flow in each link. 
From the flow, we can calculate the velocity $u_{ij} = q_{ij}/\pi r_{0,ij}^2$ 
of the fluids. We then move every bubble an amount 
$\Delta x_{ij} = u_{ij} \Delta t$, 
where the time step $\Delta t$ is chosen such that no bubble 
moves more than 10\% of the link length. 

When the fluid reaches the end of a link, it is redistributed into 
connected links in proportion to the local flow. If at any given 
point there are more than three bubbles present in a link, the closest 
two are merged such that the center of mass of the merged bubbles is conserved. 
Further details can be found in \cite{amhb98,kah02}.



\end{document}